\documentclass[useAMS,usenatbib]{mn2e}

\usepackage{latexsym}
\usepackage{color}
\usepackage{amsmath}
\usepackage{amssymb}
\usepackage{graphicx}
\usepackage{aas_macros}
\newcommand{\Ms}{M$_{\odot}$}

\title[LW Escape Fractions from the First Galaxies]{Lyman-Werner Escape Fractions from the First Galaxies}

\author[Schauer et al.]{Anna T. P. Schauer$^{1}$
\thanks{E-mail:schauer@uni-heidelberg.de},
Bhaskar Agarwal$^{1,2}$, Simon C. O. Glover$^{1}$, 
\newauthor Ralf S. Klessen$^{1}$, Muhammad A. Latif$^{3,4}$, Llu\'is Mas-Ribas$^{5}$, 
\newauthor Claes-Erik Rydberg$^{1}$, Daniel J. Whalen$^{6}$, Erik Zackrisson$^{7}$\\
$^{1}$Universit\"at Heidelberg, Zentrum f\"ur Astronomie, Institut f\"ur Theoretische
Astrophysik, Albert-Ueberle-Str. 2, 69120 Heidelberg, Germany\\
$^{2}$Department of Astronomy, 52 Hillhouse Avenue, Steinbach Hall, Yale University, 
New Haven, CT 06511, USA\\
$^{3}$Sorbonne Universites, UPMC Univ Paris 06 et CNRS, UMR 7095, Institut dAstrophysique de Paris, 
98 bis bd Arago, 75014 Paris, France\\
$^{4}$Department of Physics,  COMSATS Institute of Information Technology, Park Road, 
Islamabad, Pakistan\\
$^{5}$ Institute of Theoretical Astrophysics, University of Oslo,
P. O. Box 1029 Blindern, NO 03015 Oslo, Norway\\
$^{6}$ Institute of Cosmology and Gravitation, Portsmouth
University, Dennis Sciama Building, Portsmouth PO1 3FX, UK\\
$^{7}$ Department of Physics and Astronomy, Uppsala University, Box 515, SE-751 20 Uppsala, Sweden}

\begin{document}

%\date{Accepted 1988 December 15. Received 1988 December 14; in original form 1988 October 11}

\pagerange{\pageref{firstpage}--\pageref{lastpage}} \pubyear{2002}

\maketitle

\label{firstpage}

\begin{abstract}
Direct collapse black holes forming in
pristine, atomically-cooling haloes at $z \approx 10-20$ 
may act as the seeds of
supermassive black holes (BH) at high redshifts. 
In order to create a massive BH seed, the host halo needs to be prevented from forming stars. 
H$_2$ therefore needs to be irradiated by a large flux of Lyman-Werner (LW) UV photons 
in order to suppress H$_2$ cooling.  
A key uncertainty in this scenario is the escape fraction of LW radiation from 
first
galaxies, the dominant source of UV photons at this epoch. 
To better constrain this escape fraction, we 
have performed radiation-hydrodynamical simulations of the growth of 
H\,\textsc{ii} regions and their associated photodissociation regions in the first galaxies 
using the ZEUS-MP code. 
We find that the LW escape fraction crucially depends on the propagation of 
the ionisation front (I-front).
For an R-type I-front overrunning the halo, the LW escape fraction is always 
larger than 95\%. If the halo recombines later from the outside--in, due to a 
softened and weakened spectrum, the LW escape 
fraction in the rest-frame of the halo (the near-field) 
drops to zero.
A detailed and careful analysis is required to analyse 
slowly moving, D-type I-fronts, where the escape fraction depends on the 
microphysics and can be as small as 3\% in the near-field and 61\% in the far-field
or as large as 100\% in both the near-field and the far-field. 
\end{abstract}


% ****************************************************************************
\section{Introduction}
% ****************************************************************************
In recent years, several highly luminous quasars 
have been observed, for example a $1.2 \times 10^{10}$~\Ms\ BH at redshift $z \approx 6.3$ 
\citep{wu15} and a $9 \times 10^9$~\Ms\ BH at redshift $z \approx 7.1$ 
\citep{mort11}. These objects provide a stern challenge for current structure 
formation models 
as it is not clear yet how BHs can assemble so much mass during the 
first billion years of the Universe \citep[see a review by][]{vol10}. 

In addition, the luminous Ly$\alpha$ emitter CR7 \citep{cr7}, recently 
discovered at $z=6.6$, is found to have a unique structure. A deep HST 
image shows it consists of three clumps, separated by $\approx$ 5~kpc. 
The most luminous clump does not show any
metal lines, but the two smaller, accompanying clumps do. 
CR7 is not only very luminous, but also shows a strong He\,\textsc{ii} 1640~\AA~
line and therefore a hard spectrum. 
\citet{cr7} interpreted this object as a chemically pristine protogalaxy forming 
a large cluster of Population~III (Pop~III) stars, but other authors have 
argued that it is more likely to be a massive accreting BH
(\citealt{tilmancr7,bhaskarcr7,pcr7-15,markcr7}, but see \citealt{bowler16}).  

Explaining these observations and 
understanding how supermassive black holes (SMBHs) form and grow at high redshift
is a topic of active research. 
On the one hand, SMBH seeds have been suggested to be 
remnants from Pop~III stars \citep{mad01,fwh01,iho16}
with relatively low masses ($10^{2-3}$~\Ms) that accrete
close to the Eddington limit for large fractions of their lifetime.
On the other hand, very massive seeds ($10^{4-6}$~\Ms) could be created by the 
direct collapse of primordial gas 
(\citealt{lr94,el95,kbd04,begel06,ln06,latif13c}, see \citealt{vol10} for a review).

A necessary condition for the direct collapse scenario is a very massive 
and hot ($T > 10^4$~K) halo 
in which gas contracts quasi-isothermally without cooling and fragmenting. 
The central clump, to which the halo compresses, contains $\ge 10\%$  
of the total gas mass \citep{bl03}.
The fate of this clump is not entirely clear: it may form a massive BH 
directly, or may form a supermassive star that subsequently collapses to 
form a BH. 
To keep the collapsed gas at high temperatures, 
the gas must be metal-free \citep{osh08} and 
molecular hydrogen has to be prevented from forming or must be 
destroyed in that system, as it would otherwise cool a pristine halo 
down to temperatures of 150 K and encourage fragmentation \citep{bl03}. 
For this, a 
critical level of external LW flux is crucial 
\citep[e.g.][]{O01,sbh10,whb11,agarw12}. 

Current work (Smidt, Whalen \& Johnson, in prep.) favours the direct collapse scenario (but see 
\citealt{at14}). 
Unlike BH seeds
originating from Pop~III stars, direct collapse BH (DCBH) seeds  
are not born starving \citep{wan04}, but instead retain their 
fuel supply \citep{awa09,pm11}. 
Lower-mass BH seeds are often 
kicked out of their host halo \citep{wf12}. 
As the metal-free clump of CR7 is accompanied by two other clumps, the question arises 
whether this is an observed formation place of such a direct collapse BH. 

With this study, we impose additional constraints to the 
DCBH seed formation scenario from simulations
(see \citealt{dfm14} who show that the 
number density of DCBH seeds is highly dependent on the LW escape fraction).
The questions we address in this work are: How much LW radiation 
can actually escape an atomically cooling halo and contribute to the LW background? 
Is shielding by neutral hydrogen an important mechanism to prevent the 
built up of the LW background? 

We have recently performed a study on LW escape fractions from minihaloes
(\citealt{anna15}, from here on S15; see also \citealt{ket04}),  
finding the escape fraction varying with halo- and stellar mass from 0 to 95\%. 
We adopted two limiting cases: the near-field and the far-field. 
In the near-field case, we examine what would be seen by an observer 
in the rest-frame of the host halo. 
This is a good approximation for when the relative velocity is smaller than the
typical width of the LW lines. 
In the far-field case, we consider an observer who is significantly Doppler 
shifted with respect to the halo, either due to a large peculiar velocity 
or simply due to the Hubble flow. This is the relevant case if we are interested 
in the global build-up of the LW background. 

In this new study, we 
use our existing framework presented in S15
but focus on more massive haloes 
($10^7$ - $10^8$~\Ms), typical hosts of the first galaxies. 
Here, instead of single stars, we focus on radiation 
from stellar clusters with time-varying spectral energy distributions (SEDs). 
We consider several different star formation efficiencies (SFEs) and stellar cluster masses, 
meant to be descriptive of the first galaxies at $z \approx 10$.

Our paper is structured as follows. In Section 2, we briefly summarize the 
semi-analytic methods described in S15 and introduce our numerical 
simulations. Details of our parameter space, 
including the choice of haloes, SEDs and SFEs are described in Section 3. 
We present our results in Section 4, where we show the dependence of 
the LW escape fractions for all halo, SED and SFE combinations over time in both 
near- and far-field. In addition, we provide tabulated results with values 
averaged over the lifetime of stellar populations. Finally, we discuss 
our findings and conclude in Section 5.

% ***************************************************************************
\section{Methods}
% ***************************************************************************
In this section, we describe the methods used to calculate the LW 
escape fractions in the near- and far-field limits.
The influence of radiation from a central stellar cluster on the gas in a 
surrounding protogalaxy is modelled in 1D using 
the ZEUS-MP hydrodynamical code (Section 2.1).
The simulation results are then post-processed to derive the LW escape fraction
(Section 2.2). 

\subsection{ZEUS-MP}
Our simulations are run with the radiation-hydrodynamics code 
ZEUS-MP \citep{wn06}. 
A full description of the code can be found in more detail in S15. Here, 
we only give a short overview of the most important features. 

The version of ZEUS-MP used in this study couples photon-conserving ionising 
UV transport self consistently to non-equilibrium primordial chemistry and 
hydrodynamics to model the growth of cosmological ionisation fronts \citep[I-fronts;][]{wn06}.
We include a network of nine chemical species (H, H$^{+}$, He, He$^{+}$, 
He$^{2+}$, H$^{-}$, H$^{+}_{2}$, H$_{2}$ and e$^{-}$). Heating and cooling processes 
like photoionisation, collisional ionisation, excitation and recombination, 
inverse Compton scattering from the CMB, bremsstrahlung emission and H$_2$ cooling
mostly following the implementation in \citet{anet97}. Details of the radiative 
reactions can be found in Table~1 of \citet{wn08b}. 

The spectra are tabulated as a function of time, in 120 energy bins with 
40 uniform bins from 0.755 to 13.6 eV and 80 uniform logarithmically spaced bins 
from 13.6 eV to 90.0 eV.
Since single stars were analysed in S15, we used SEDs that were constant in time.
However, in this work as we are focusing on stellar populations, 
we consider time-varying SEDs as described in Section 3.2.

For the simulation, we assume spherical symmetry. The volume is divided into 1000 radial
cells, reaching out to about twice the virial radius. 
We use a logarithmically spaced grid, allowing us to have high resolution within the inner
regions of the halo without compromising the running time. 
For 8 out of a total of 30 
simulations, we additionally refine a 20~pc region around the I-front with 20000 cells 
to achieve numerical
convergence. 
For all setups, we run comparison simulations with half the number of cells. In all cases, 
the results from the comparison simulations differ by less than 1\% from the original
simulations, demonstrating that our results have converged numerically. 

\subsection{Semi-analytical post-processing}
After the simulations have run, 
we calculate escape fractions of LW photons 
in two limiting cases, the near-field (Section 2.2.1) and the far-field 
(Section 2.2.2). For a detailed discussion of these two limits, 
we refer the reader to S15. We summarize the calculations in the sections below. 

\subsubsection{Near-field}
The near-field limit applies in the vicinity of the halo and 
is the escape fraction that would be measured by an observer who is at rest 
with respect to the halo, or moving with only a low velocity relative to it. 
In this limit, H$_2$ within the halo can potentially provide effective 
self-shielding of the LW lines. 

\citet{db96} and \citet[][hereafter WH11]{wh11}, have calculated the 
effects of H$_2$ self-shielding in the low density and LTE limits, 
respectively. They provide simple fitting functions describing the shielding 
as a function of H$_2$ column density and temperature. 
LW photons that do not dissociate H$_2$ molecules 
or are not absorbed and re-emitted as lower-energy photons, can escape the halo. 
For high column densities, H$_2$ can shield itself against LW radiation, or can 
be shielded by H. We can therefore equate the escape fraction to the shielding 
fraction of molecular, or molecular and neutral hydrogen. 

For the optically thick case ($N_\mathrm{H2} > 10^{13}$~cm$^{-2}$), 
we use the shielding functions from WH11, which depend on 
the column density $N_\mathrm{H2}$ and, via the Doppler width, on temperature and 
velocity dispersion of the gas in the halo. 
We consider two cases: one in which only H$_2$ contributes, and another where we 
additionally account for shielding from neutral hydrogen. In the H$_2$-only case, 
we use the shielding functions given in Equation (8) of WH11. When we account 
for shielding from atomic hydrogen, we additionally use 
equations (11) and (12) from the same paper. 
For small column densities ($N_\mathrm{H2} < 10^{13}$~cm$^{-2}$), 
we set the escape fraction to~1.

\subsubsection{Far-field}
The far-field 
limit corresponds to the escape fraction that would be measured by an observer located 
far away from the halo, moving with a large peculiar velocity or Hubble 
flow velocity with respect to it. 
In this limit, the LW lines 
in the restframe of the observer
are Doppler shifted and do not coincide with the LW lines in the halo. 

To calculate the escape fraction in this limit,  we choose the LW range 11.2 -- 13.6 eV
(corresponding to a wavelength range of 912 -- 1110~\AA) and calculate what
fraction of the light in this range will be absorbed by H$_2$ or H in the halo. 
If we denote this fraction as $f_\mathrm{abs}$, then the far-field escape fraction is 
simply $f_\mathrm{esc} = 1$ - $f_\mathrm{abs}$. 

To calculate $f_\mathrm{esc}$, we first compute the dimensionless 
equivalent width of the full set of accessible LW lines. To do this, 
we follow \citet{rw74} and calculate the equivalent width of every line, 
accounting for both Lorentz broadening and Doppler broadening. 

We then sum the equivalent widths of the individual lines and add to them the 
dimensionless equivalent width of all Lyman series transitions 
of atomic hydrogen 
that lie within 
the energy range of interest, including transitions up to $n=10$. 
\citet{ar89} and \citet{abg92} provide data on 
the total radiative de-excitation rates of the excited states,  
oscillator strengths and transition frequencies of molecular hydrogen.
\citet{wf09} give simulated data for atomic hydrogen. 
In a final step, we combine the dimensionless equivalent widths and 
follow the prescription in \citet{db96}, and divide by the total width of the 
LW band. This yields $f_\mathrm{abs}$, from which $f_\mathrm{esc}$  
follows trivially.
% ***************************************************************************
\section{Parameter Space}
% ***************************************************************************
In this section, we present our parameter space. 
We consider two haloes with masses of $5.6\times 10^7$ (Halo A) 
and $4\times 10^8$~\Ms (Halo B), 
three SFEs and five SEDs that evolve over 
time for a given total stellar mass in the halo.  The haloes in our study are 
taken from cosmological simulations by \citet{lv15}.  
\begin{table}
\begin{center}
\begin{tabular}{cccc}
$M_\mathrm{min}$ [\Ms] & $M_\mathrm{max}$ [\Ms] & runtime [Myr] & shape \\
\hline
50  & 500 & 3.6  & Salpeter \\
9   & 500 & 20.2 & flat \\
1   & 500 & 20.2 & log-normal \\
0.1 & 100 & 20.2 & Kroupa   \\
0.1 & 100 & 20.2 & Kroupa + nebula \\
\end{tabular}
\caption{Summary of all spectral energy distributions used in this study.}
\label{tab:sed}
\end{center} 
\end{table}
\subsection{Haloes}
\begin{figure*}
\includegraphics[width=1.99\columnwidth]{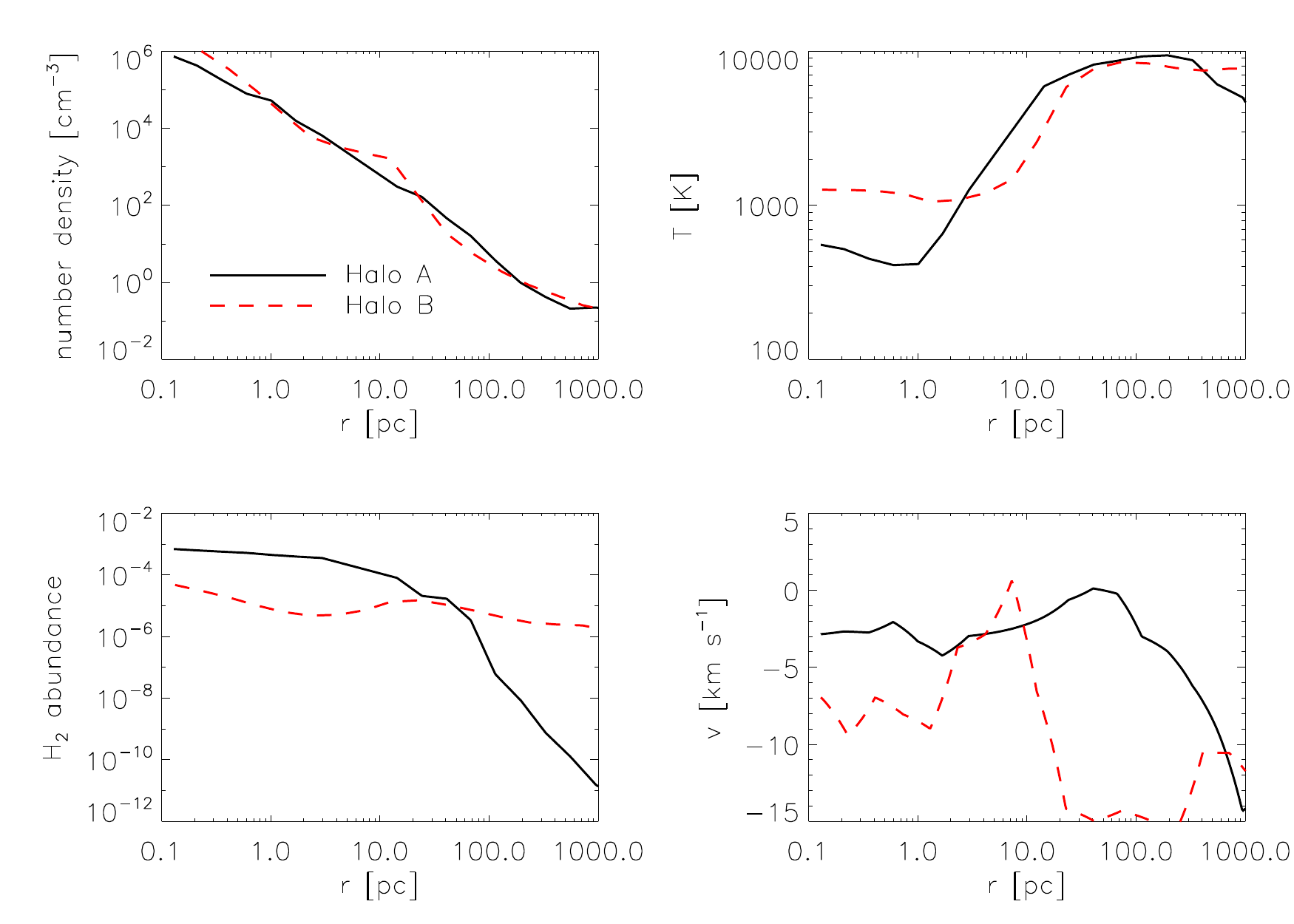}
\caption{Number density (upper left panel), temperature (upper right panel), 
H$_{2}$ abundance (lower left panel) 
and velocity (lower right panel) profiles at startup. Halo A (black solid line) has a total 
mass of $5.6\times 10^7$~\Ms, Halo B (red dashed line) 
has a total mass of $4\times 10^8$~\Ms. }
\label{fig:haloes}
\end{figure*}
We perform our simulations with data from cosmological simulations from \citet{lv15}. 
All three dimensional quantities are mapped onto one dimensional radial profiles. 
In Halo A, we smooth out a region at radius~$\approx$~20~pc, where a substructure is falling 
into the halo. In our one dimensional picture, this provides a more physical result. 
In Figure~\ref{fig:haloes}, we show the temperature, density and velocity profiles as well 
as the initial H$_{2}$ abundance. Both haloes follow a similar density profile, but 
Halo A shows a higher abundance of molecular hydrogen and therefore lower temperatures in 
the centre. 
Both haloes are selected at redshift $z = 10$ and assumed to be metal free. The virial radii 
of the objects are 1.3 and 4.3~physical kpc, respectively.
\subsection{Spectral energy distribution}
\begin{figure}
\includegraphics[width=0.99\columnwidth]{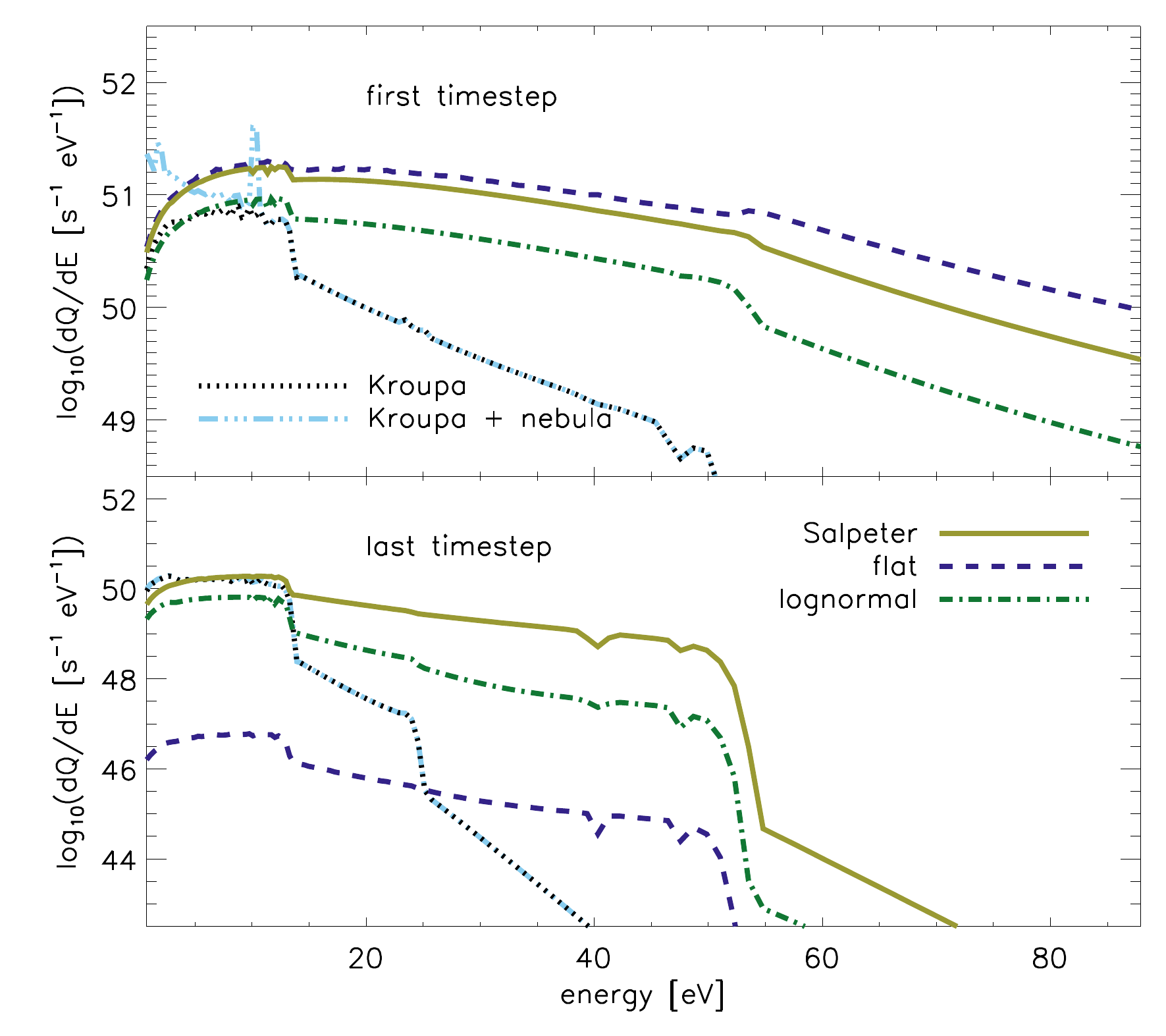}
\caption{SEDs for a SFE of 0.5\% in Halo A. 
Different colours and linestyles mark different SEDs: 
a Salpeter IMF (olive solid line), 
a flat IMF (dark blue dashed line), 
a lognormal IMF (green dot-dashed line), 
a Kroupa IMF without nebular emission (black dotted line) 
and a Kroupa IMF with nebular emission (light blue dot-dot-dot-dashed line). 
The upper panel shows the initial spectrum, 
the lower panel, shows the 
final spectrum at the cut-off of the simulation at 
3.6 and 20.2~Myr, for the Salpeter and other IMFs, respectively. }
\label{fig:sed}
\end{figure}

\begin{figure}
\includegraphics[width=0.99\columnwidth]{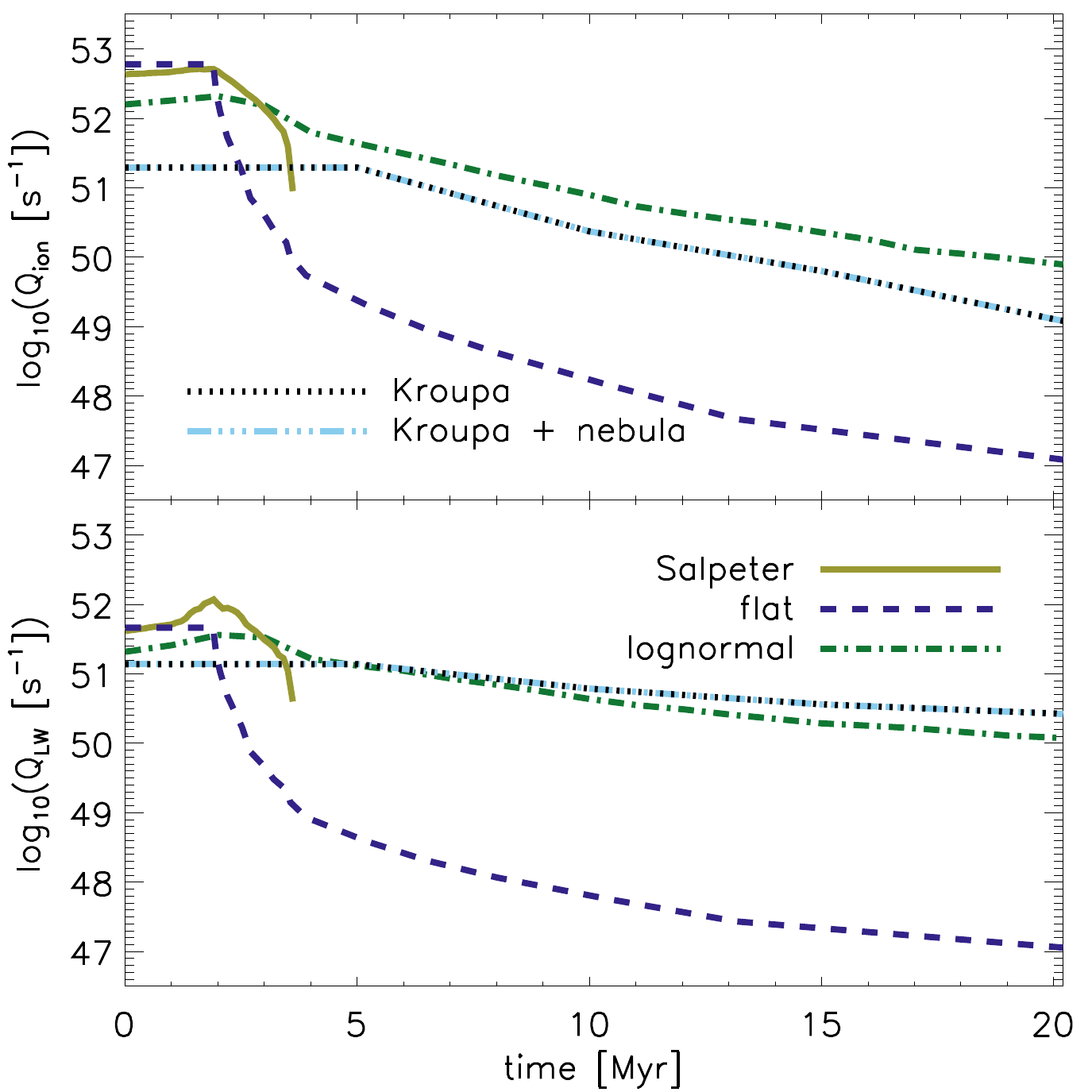}
\caption{Ionising (upper panel) and Lyman-Werner (lower panel) photon production rates for 
all SEDs for a SFE of 0.5\% in Halo A. The colour-coding is the same 
as in Figure~\ref{fig:sed}.}
\label{fig:photoncounts}
\end{figure}

We study SEDs derived from direct sums of individual Pop~III stars and from spectral
synthesis models of Pop~III and Pop~III / Pop~II stellar populations calculated 
with the Yggdrasil
and Starburst99 codes.  
As the IMF of first stars is still very uncertain 
\citep{cgk08,stacy10,cgkb11,sb14,hir14,stacy16}, we
consider different descriptions that cover the range currently discussed in the 
literature. A total of 5 initial mass functions (IMFs) for the stars 
are considered. We use different methods for creating the time-dependent SEDs; 
for a flat IMF (9 -- 500~\Ms) we use spectra from Pop~III stars alone, 
for a Salpeter slope IMF (50 -- 500~\Ms) and a lognormal IMF (1 -- 500~\Ms), 
we make use of the code Yggdrasil and for Kroupa IMFs (0.1 -- 100~\Ms), 
we use a combination of Yggdrasil and Starburst 99; all details can be 
found in the following subsections. 
For all SEDs, we assume an instantaneous starburst at the beginning. When 
the runtime exceeds the lifetime of a star, that star does not explode in a supernova, 
but is simply removed from the spectrum. We discuss the validity of this assumption 
in Section 4.3. 
We show the SED for our fiducial halo-SFE case in Figure~\ref{fig:sed} 
for the first and last moment of the simulation for all five SEDs in 
terms of $\mathrm{d}Q/\mathrm{d}E$, 
where $\mathrm{d}Q/\mathrm{d}E$ is the number of photons emitted per second and per eV. 
Depending on the SED, the simulation runtimes differ. The SED 
with a Salpeter slope ranging from 50 -- 500~\Ms\ lasts for 3.6~Myr, 
corresponding to the lifetime of the least massive star. 
For all other SEDs, we adopt a runtime given by the lifetime of a 9~\Ms\ star, 20.2~Myr 
\citep{schae02}. A summary 
of the SEDs can be seen in Figure~\ref{fig:photoncounts}, where we show
the number of ionising photons and LW photons produced as a function 
of time. 
\subsubsection{Flat slope IMF from Pop~III Stellar Spectra}
We create a stellar population from a flat slope IMF within the mass range
9 -- 500~\Ms, using the stellar SEDs and same methodology as in
\citet{lluis16}; we refer the reader to that work for a detailed
description of the calculations. These authors used parameters from 
\citet{schae02} and \cite{mar01}, and assumed the stars to be metal-free and
to reside on the main sequence. We account for the stellar evolution over time
by simply removing from the calculation the stars that reach the end of their
lives. This SED is considered a radiation source until the least massive star
disappears, t$_\mathrm{end}$ = t$_{9\mathrm{M}_\odot}$ = 20.2~Myr.
\subsubsection{Salpeter and lognormal IMF created with Yggdrasil}
To generate SEDs for Pop~III star clusters with the Yggdrasil spectral synthesis code, we sum
the spectra of its constituent stars \citep{zack11,ry15}.  The cluster is assumed to 
form instantaneously. In summing the spectra of individual stars to obtain an 
SED we consider two top-heavy IMFs.
One is a Salpeter-like extremely top-heavy IMF ranging from 50 -- 500~\Ms\ with a typical mass
of 100~\Ms\ \citep{schae02}.  The second is a log-normal distribution IMF with a 
typical mass of 10~\Ms\ and the standard deviation of the underlying normal distribution 
$\sigma=1.0$, ranging from 1~\Ms\ to 500~\Ms\ (\citealt{raiter10}, see \citealt{tum06}).  
Yggdrasil provides
spectra for star clusters for ages of up to $3.6$~Myr for the extremely top-heavy IMF, which is
the lifetime of the longest-lived star in the cluster. 
The log-normal distribution is turned off after 20.2~Myr. 
\subsubsection{Kroupa IMF with and without nebular emission created with Yggdrasil / Starburst99}
The purely stellar SEDs are based on synthetic spectra from \citet{raiter10} 
for absolute metallicities $Z=10^{-7}$ and $Z=10^{-5}$ and Starburst99 \citep{starburst99} 
with Geneva high mass-loss tracks in the case of $Z=0.001$, 0.004, 0.008 and 0.020. 
Nebular emission has been added using the Cloudy photoionisation code 
\citep{cloudy13} with parameters as in \citet{zack11}, under the 
assumption of no Lyman continuum leakage and no dust. The stellar SEDs have 
been rescaled in all cases to 
be consistent with the \citet{kroupa01} stellar 
initial mass function.

We point out here that there are some physical limitation of the Kroupa plus nebular 
emission case. In principle, the radiation in the halo would be a mixture of 
starlight and nebular emission. No datasets of mixed SED grids are available 
at this time, and we simply add the nebular to the stellar emission. Thus we 
urge the reader to treat this case as an extreme limit where the nebular 
emission has been maximised. Excluding the Kroupa plus nebular emission case 
from our work does not qualitatively change our results. 
\subsection{Star Formation Efficiencies}
The SEDs from the previous section are normalised to a total number of stars, 
calculated for each halo. For the fiducial case, we adopt the stellar to virial
mass ratio from \citet{oshea15} 
\begin{equation}
f_\star = 1.26\times 10^{-3} \left(M_\mathrm{vir} / 10^8 M_\odot \right)^{0.74} ,
\end{equation}
which is valid in the mass range $10^7$~\Ms $\le M_\mathrm{vir} \le 10^{8.5}$~\Ms. Our most massive 
halo, Halo B with $M_\mathrm{vir} = 10^{8.6}$~\Ms, is slightly outside this range
and we overestimate the stellar mass by a small amount. 

The stellar mass, $M_\star = f_\star \times M_\mathrm{vir}$, relates to a 
star formation efficiency: 
\begin{equation}
\mathrm{SFE} = \frac{M_\star}{M_\mathrm{b}} ,
\end{equation}
where $M_\mathrm{b}$ is the baryonic mass in a halo approximated by 
$M_\mathrm{b} = \Omega_\mathrm{b} / \Omega_0 \times M_\mathrm{vir}$. 
Using the values from \citet{wmap7b}, $\Omega_\mathrm{b}=0.0456$ and 
$\Omega_0=0.2726$, we derive SFEs of 0.5\% (Halo A) and 2.1\% (Halo B). 
In addition, we adopt an upper and a lower SFE limit on each halo, ranging from 0.1\%
to 5.0\%. All properties are listed in Table \ref{tab:sfe}. 

For our lower limit SFE of 0.1\% in Halo A we get a total of $\approx 10^4$~\Ms~in 
stars. \citet{lluis16} show that for a total stellar mass of $10^3$~\Ms, there 
is a factor of few difference in the LW flux for different stochastically sampled 
IMFs, but this is much reduced for $\approx 10^4$~\Ms~in stars. 

\begin{table*}
\begin{center}
\begin{tabular}{lccccc}
 & $M_\mathrm{vir}$ [\Ms] & $M_\mathrm{b}$ [\Ms] & $M_{\star}$ [\Ms] & $f_\star$ [\%] & SFE [\%] \\
\hline
Halo A, fiducial    & $5.6 \times 10^7$ & $9.4 \times 10^6$ & $4.6 \times 10^4$ & 0.082 & 0.5 \\
Halo A, lower limit &                   &                   & $9.4 \times 10^3$ & 0.015 & 0.1  \\
Halo A, upper limit &                   &                   & $9.4 \times 10^4$ & 0.15  & 1.0  \\
\hline
Halo B, fiducial    & $4.0 \times 10^8$ & $6.7 \times 10^7$ & $1.4 \times 10^6$ & 0.35  & 2.1  \\
Halo B, lower limit &                   &                   & $6.7 \times 10^5$ & 0.16  & 1.0  \\
Halo B, upper limit &                   &                   & $3.3 \times 10^6$ & 0.79  & 5.0  \\
\end{tabular}
\caption{Star formation efficiencies and the corresponding stellar masses for 
both haloes. The fiducial SFEs are calculated using a formula relating 
the mass in stars to the total mass from \citet{oshea15}; see Equation~1.}
\label{tab:sfe}
\end{center} 
\end{table*}
%

% ***************************************************************************
\section{Results}
% ***************************************************************************
Consistent with S15, we find that 
the LW escape fraction depends on the evolution of the ionisation 
front (I-front). In a completely ionised halo, the escape fraction is 100\%, 
but when the I-front retreats or only slowly expands, molecular hydrogen 
production is enhanced, allowing it to self-shield.
The propagation of the I-front is therefore an important quantity 
for the calculation of the escape fraction. 
We identify three different ways the I-front can behave and 
discuss one prominent example of each type in 
detail in Section 4.1. In the following subsections we will tabulate 
the time-averaged escape fractions in both near-field (Section 4.2)  
and far-field (Section 4.4) as well as the escape fraction averaged over the 
LW flux (Section 4.3). 
We also present estimates for the ionising escape fractions based on the position
of the I-front in Section 4.5. The escape fractions we provide are lower limits: 
our one dimensional simulations cannot capture 
asymmetric break-out along directions of minimum column density and we 
expect radiation to break out faster in some low-density regions. 
%
%%%
\subsection{Individual ionisation front behaviour}
%%%
%
\subsubsection{Outbreaking ionisation front}
\begin{figure}
\includegraphics[width=0.99\columnwidth]{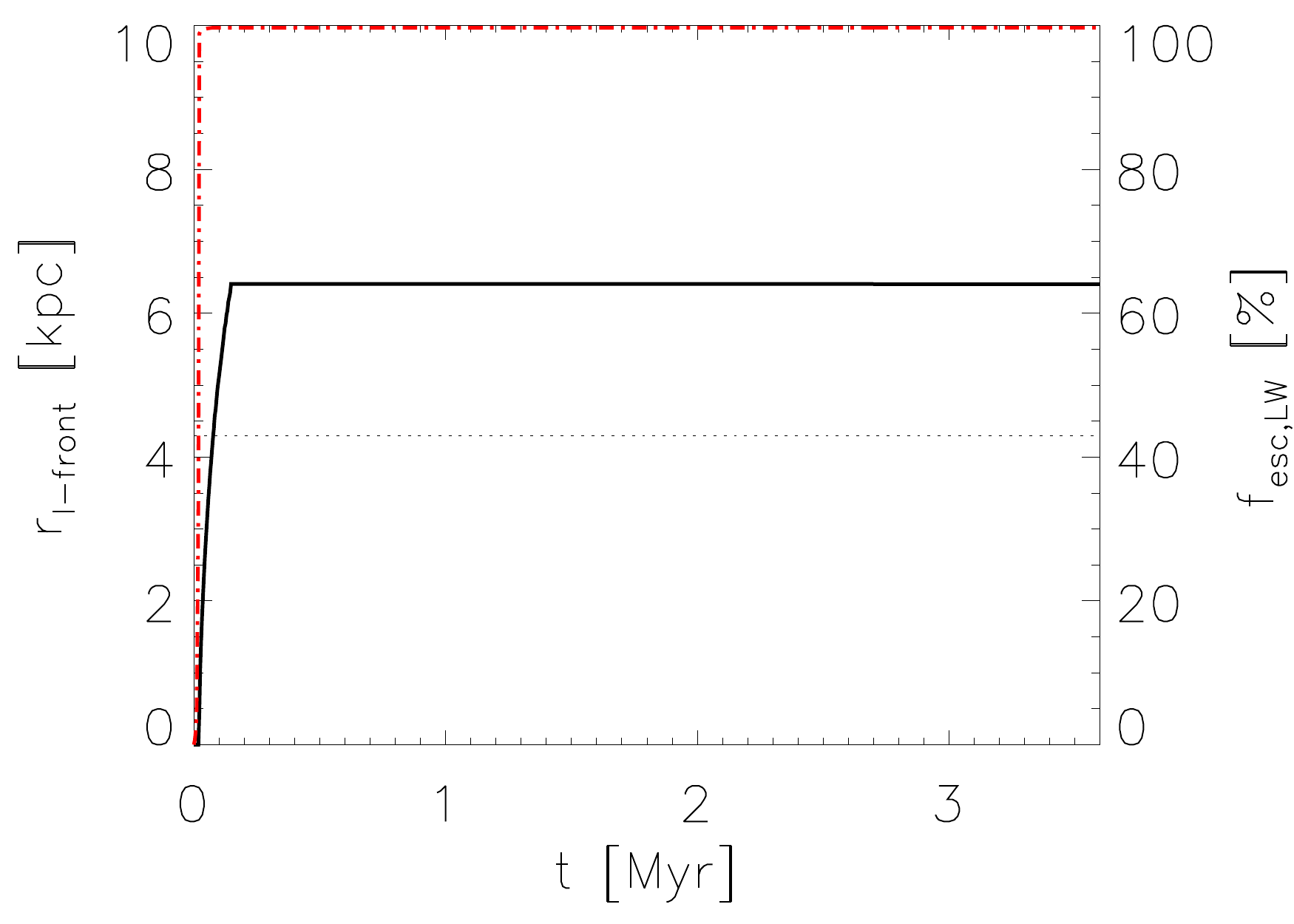}
\caption{Position of the I-front (black solid line; left axis) and LW escape fraction in the near-field 
(red dot-dashed line; right axis) as a function of time for a 1.0\% SFE with a Salpter IMF in Halo B. 
The virial radius is shown by a black dotted line. }
\label{fig:ifront-busal100}
\end{figure}
\begin{figure*}
  \includegraphics[width=0.99\columnwidth]{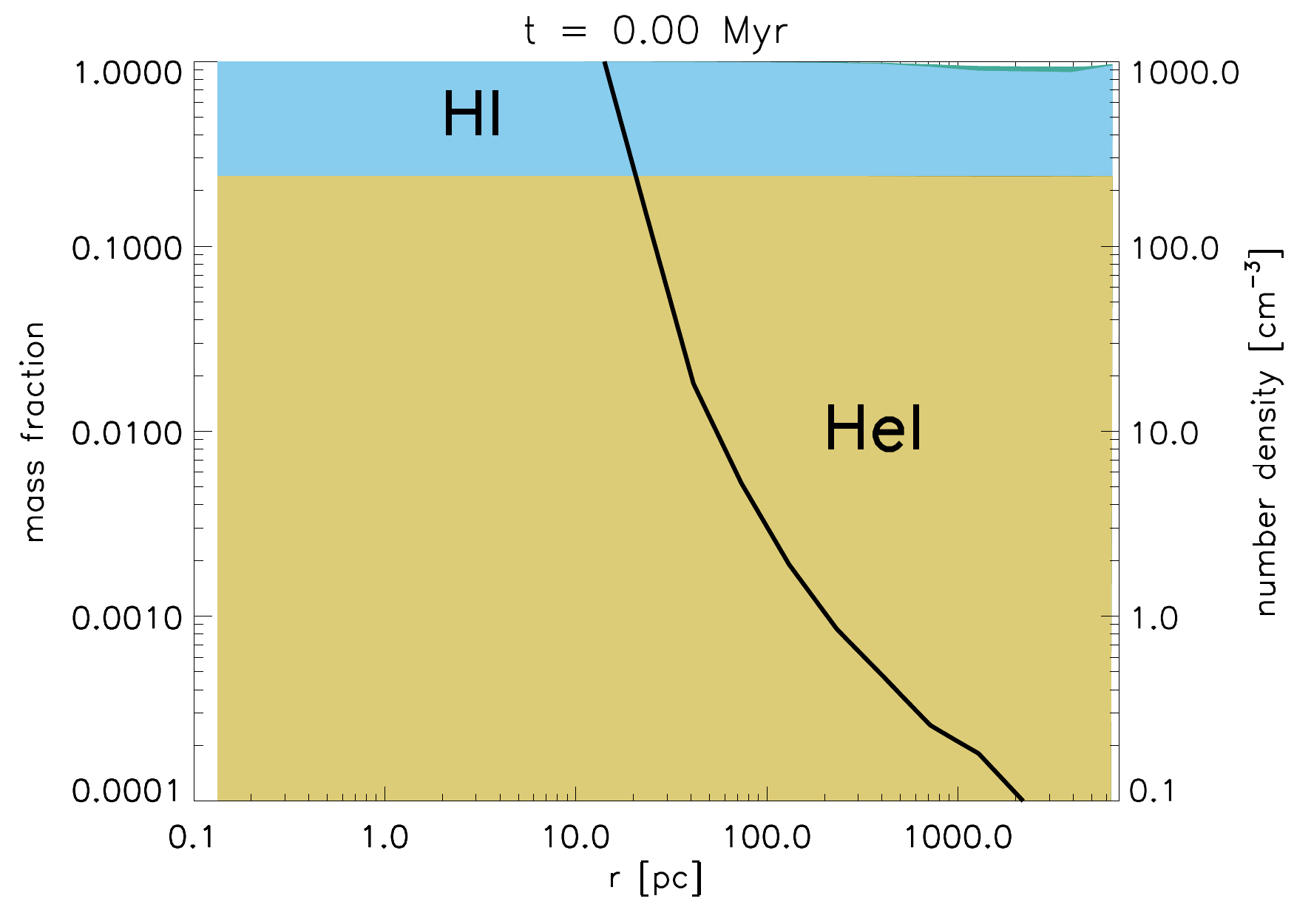}
  \includegraphics[width=0.99\columnwidth]{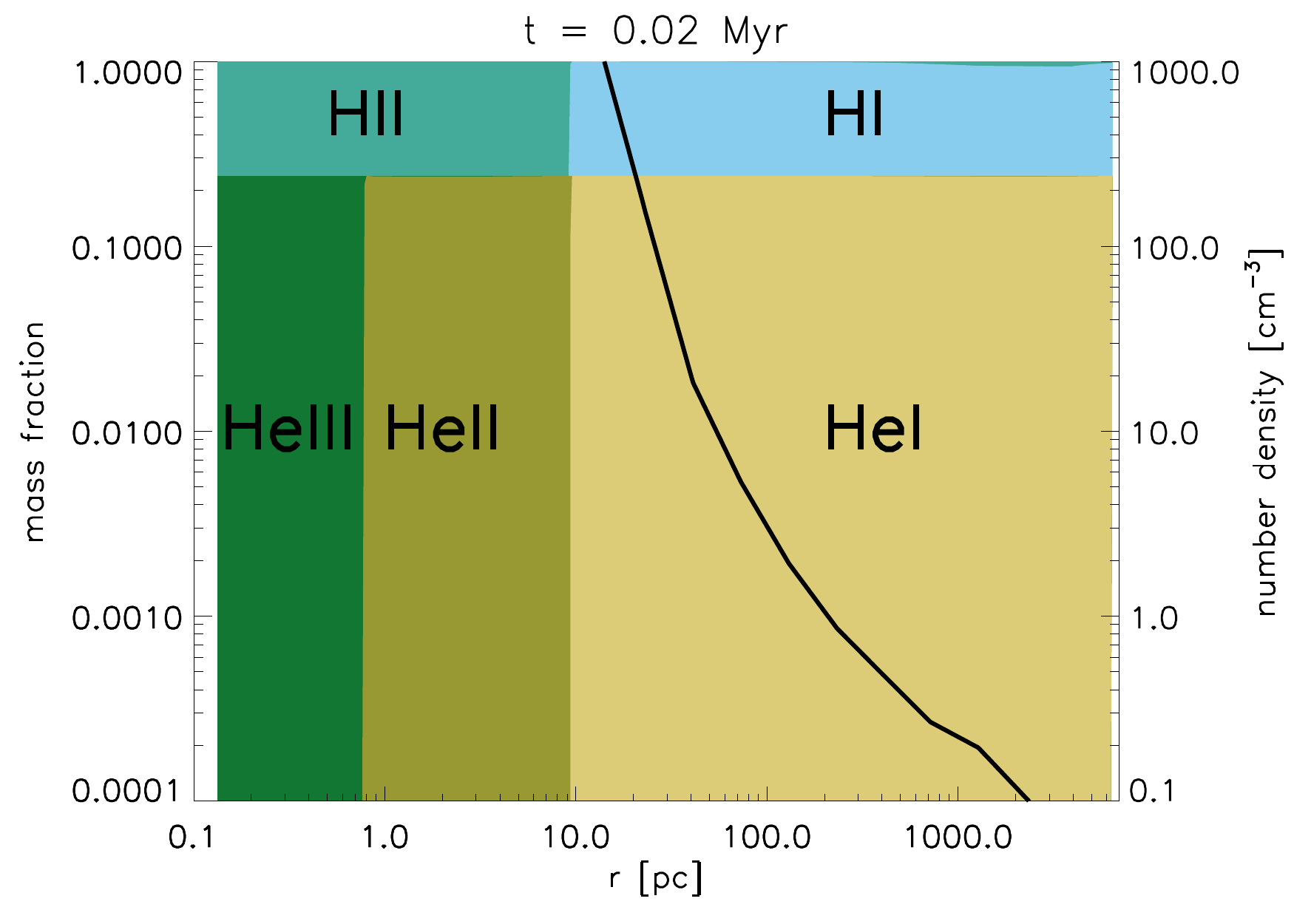}
  \includegraphics[width=0.99\columnwidth]{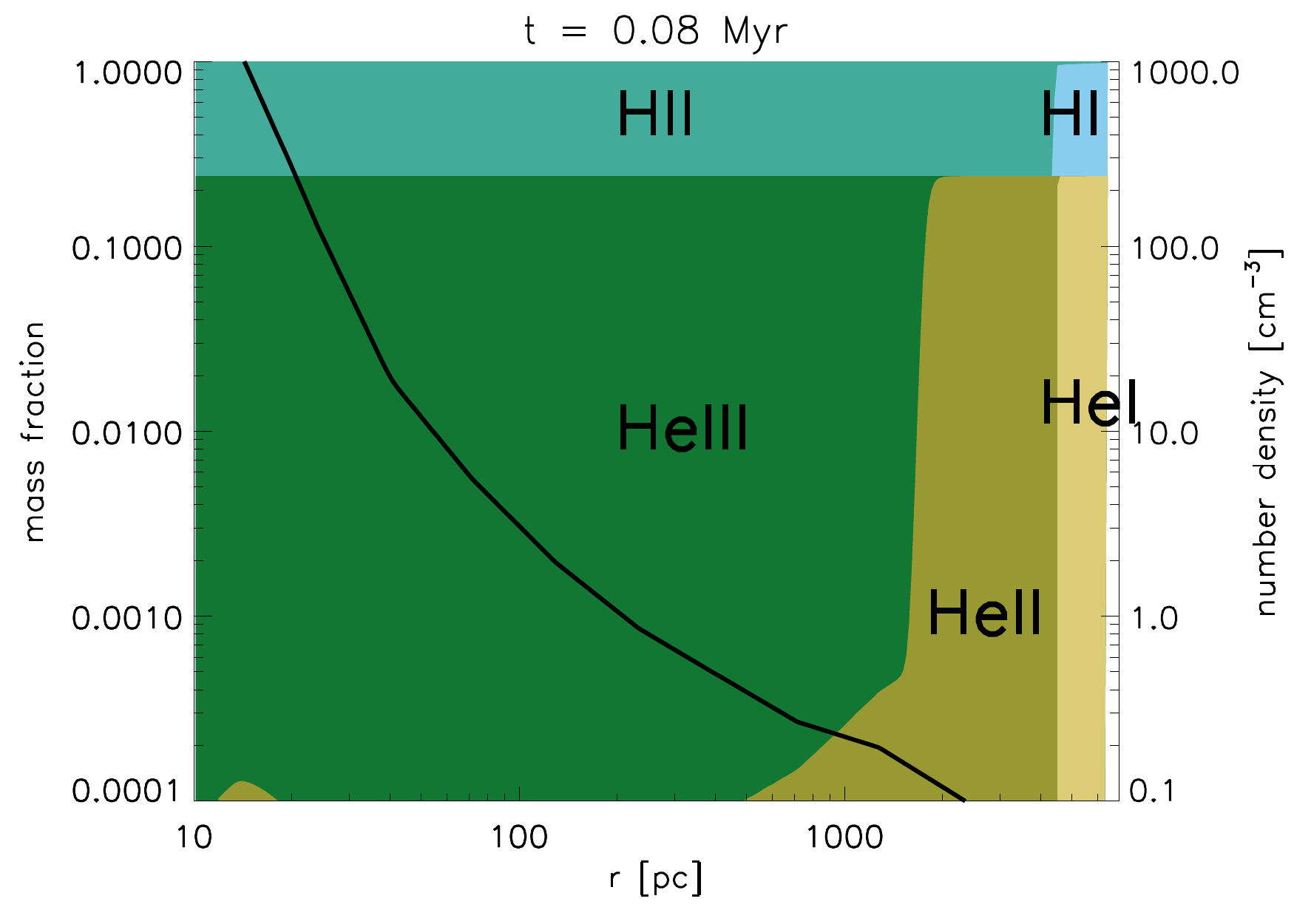}
  \includegraphics[width=0.99\columnwidth]{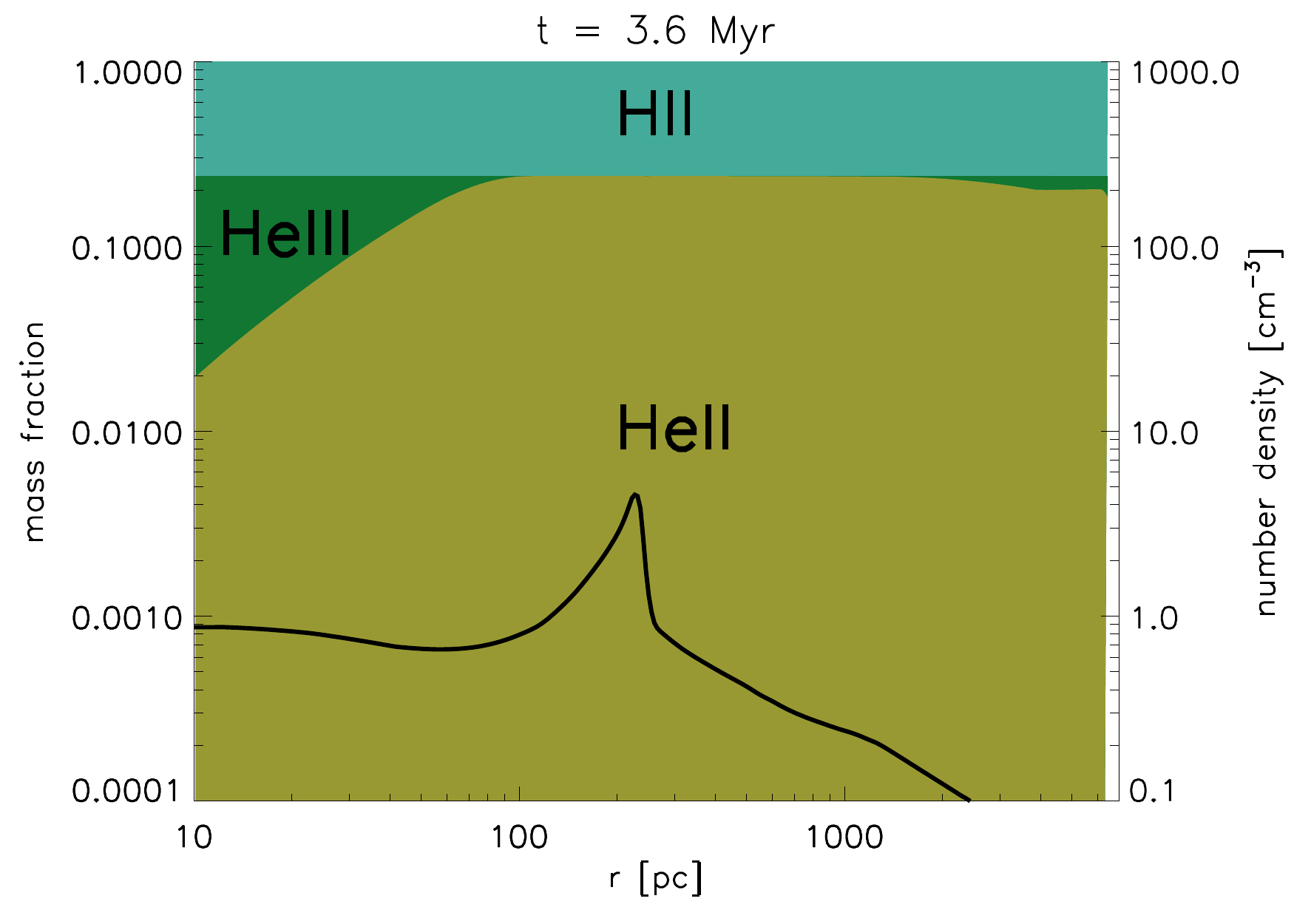}
\caption{Abundances of the primordial species as a function of radius for four output times for a 
1.0\% SFE with a Salpter IMF in Halo B. The black solid line shows the number
density of the gas against radius (corresponding to the
label on the right side of the plots).} 
\label{fig:abund-busal100}
\end{figure*}
In the case of a strong radiation source, the ionisation front quickly breaks 
out of the halo. Simultaneously, 
the escape fraction jumps up to 100\% because there is no molecular hydrogen within 
the ionised region. The lower limit case SFE (1\%) for our Salpeter SED 
in Halo B provides a prominent example. Figure~\ref{fig:ifront-busal100} displays the 
position of the I-front and the LW escape fraction in the near-field as a function of time. 

The abundances of the main chemical species can be seen in Figure~\ref{fig:abund-busal100} 
for four different output times. All species are colour coded. 
The black solid line shows the number density in cm$^{-3}$. 
Initially (upper left panel), the halo is mostly neutral with only a small amount of ionised 
hydrogen outside 200~pc. Already at 0.02~Myr (upper right panel), a very steep I-front has advanced out to 9~pc, 
indicated by the turquoise H\,\textsc{ii} region that has a sharp transition to the light blue 
neutral H outer region. The I-front continues to proceed as an R-type I-front through the halo and crosses the virial
radius at 0.08~Myr (lower left panel). Helium is doubly ionised out to 1000~pc. The density 
still shows its initial profile. At the end of the simulation, 
at 3.6~Myr (lower right panel), the halo 
is still completely ionised.
The expansion of the ionised gas has started to drive a shock front outwards, but this has 
only reached 200~pc. It is therefore well within the H\,\textsc{ii} region and hence is 
not relevant for our LW escape fraction calculation. 
Throughout the simulation, the H$_2$ abundance remains 
much smaller than $10^{-4}$ and the halo is optically 
thin to LW radiation. The time-averaged LW escape fraction in the near-field therefore is 
99\% in this case. In general, 
\begin{equation} 
f_{\mathrm{esc,LW}}^{\mathrm{(o)}} \ge 0.95  
\end{equation}
for both the near-field and far-field limit 
in all models with I-front breakout (indicated by the superscript $^{\mathrm{(o)}}$). 
\subsubsection{Outbreaking and returning ionisation front}
\begin{figure}
\includegraphics[width=0.99\columnwidth]{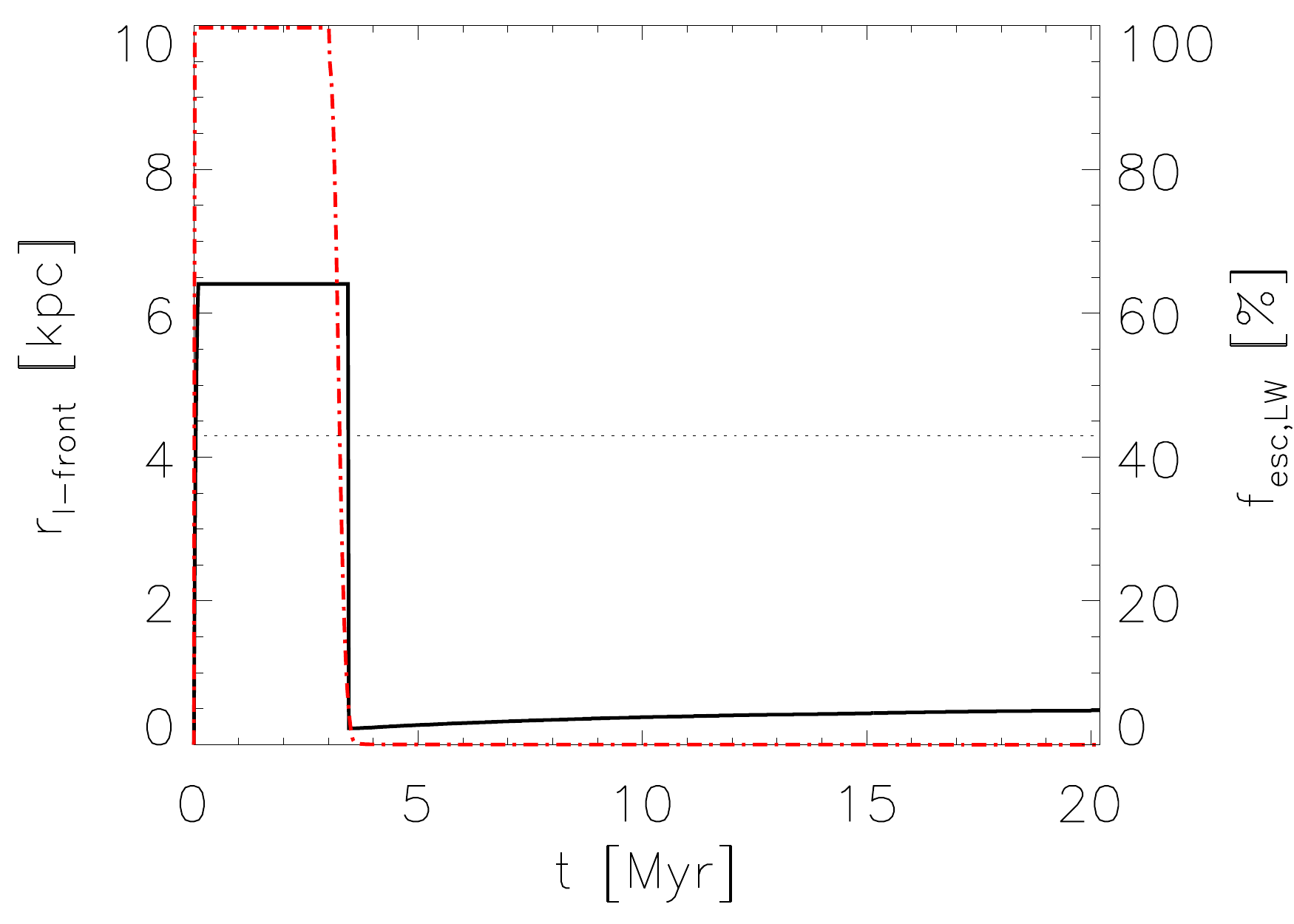}
\caption{Position of the I-front (black solid line) and LW escape fraction in the near-field 
(red dot-dashed line) as a function of time for a 1.0\% SFE with a flat IMF in Halo B. }
\label{fig:ifront-blluis100}
\end{figure}
\begin{figure*}
  \includegraphics[width=0.99\columnwidth]{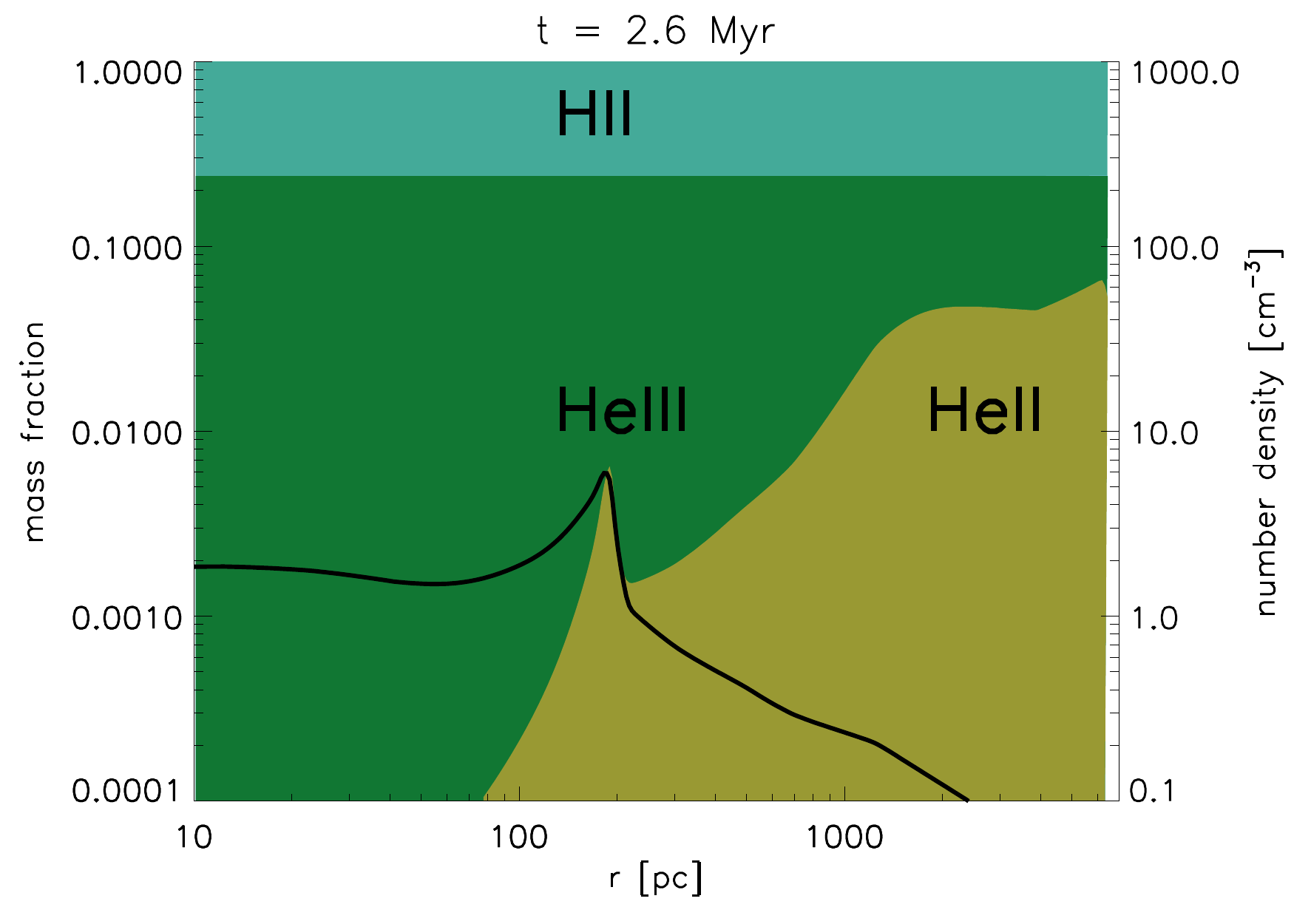}
  \includegraphics[width=0.99\columnwidth]{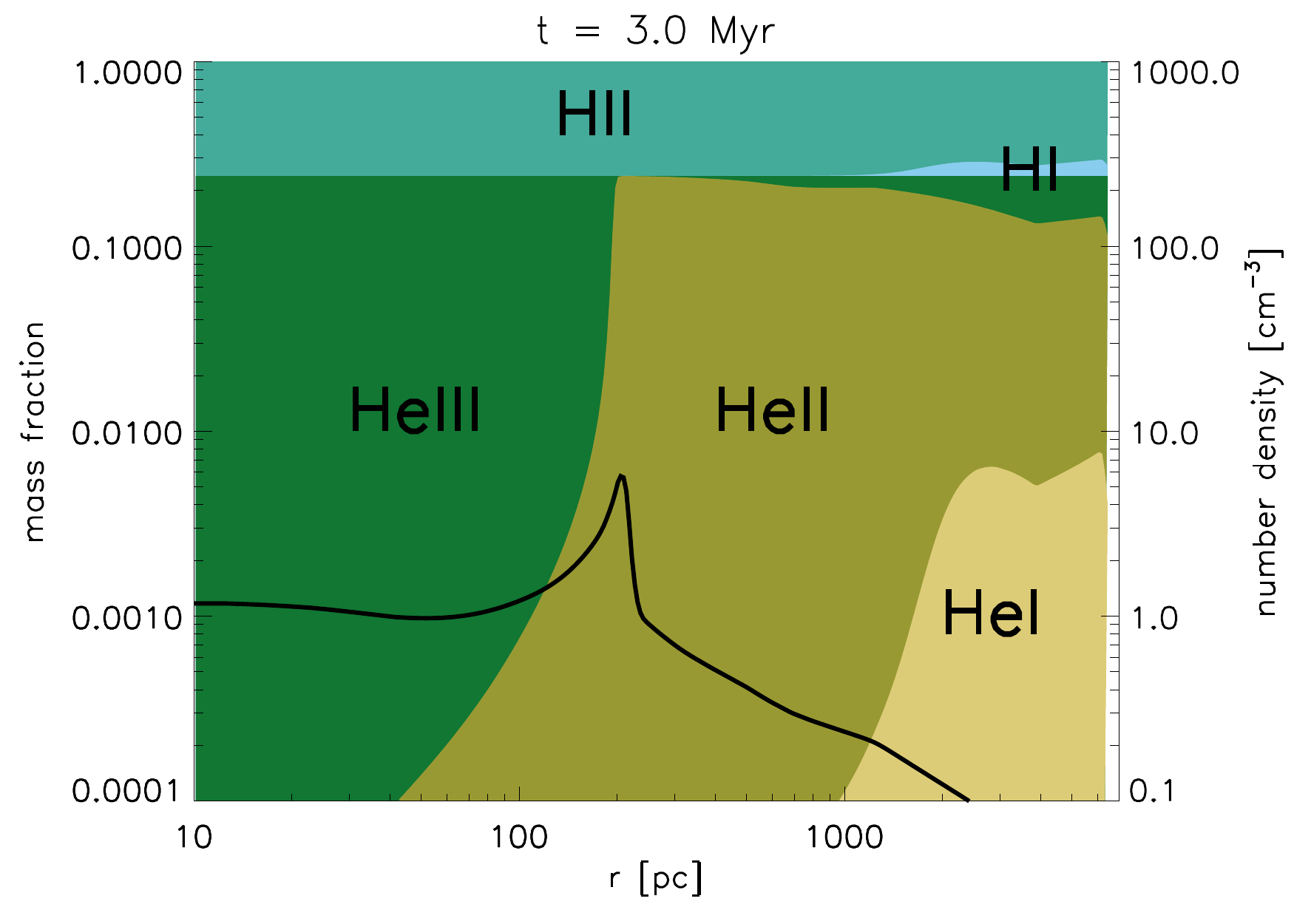}
  \includegraphics[width=0.99\columnwidth]{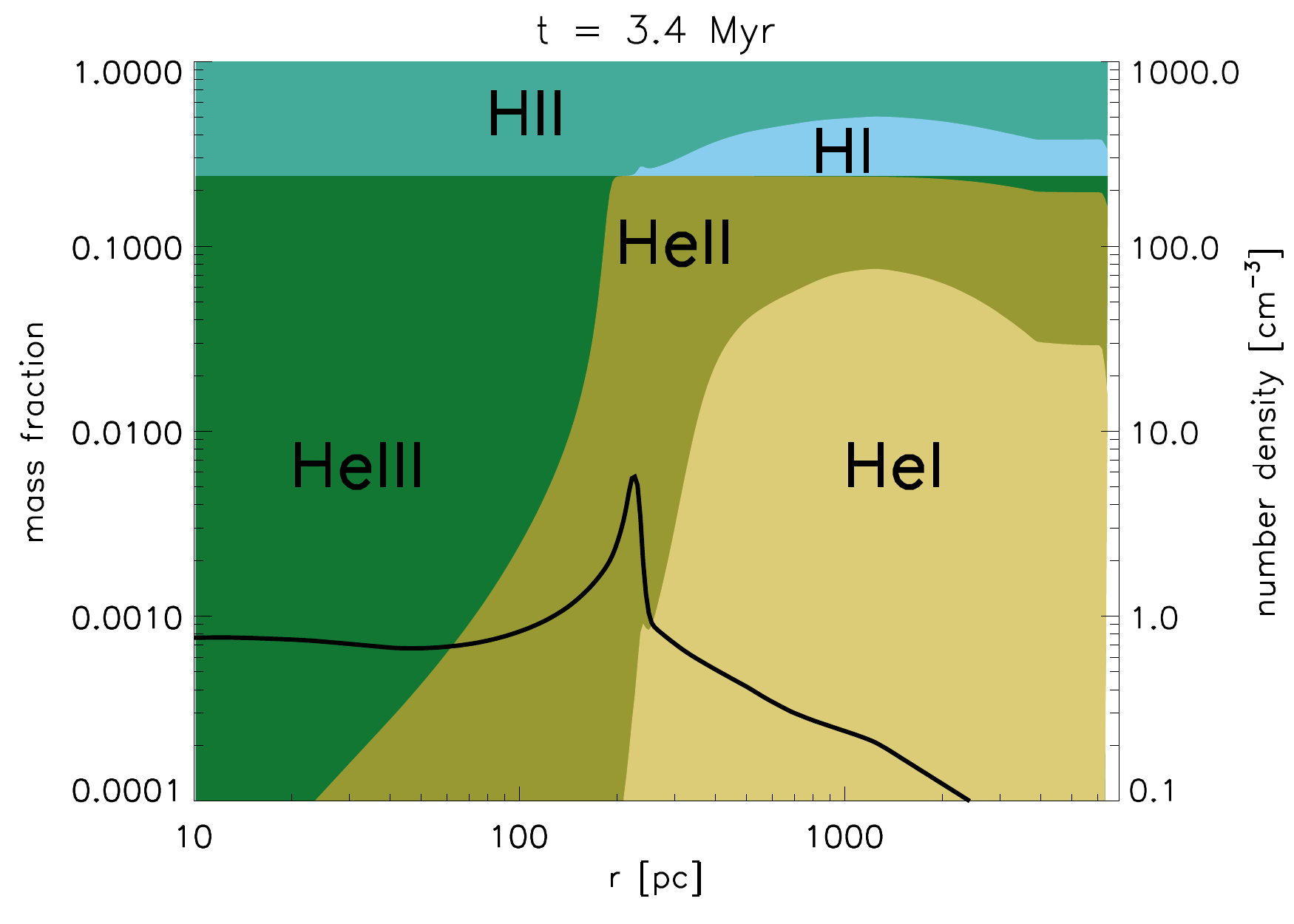}
  \includegraphics[width=0.99\columnwidth]{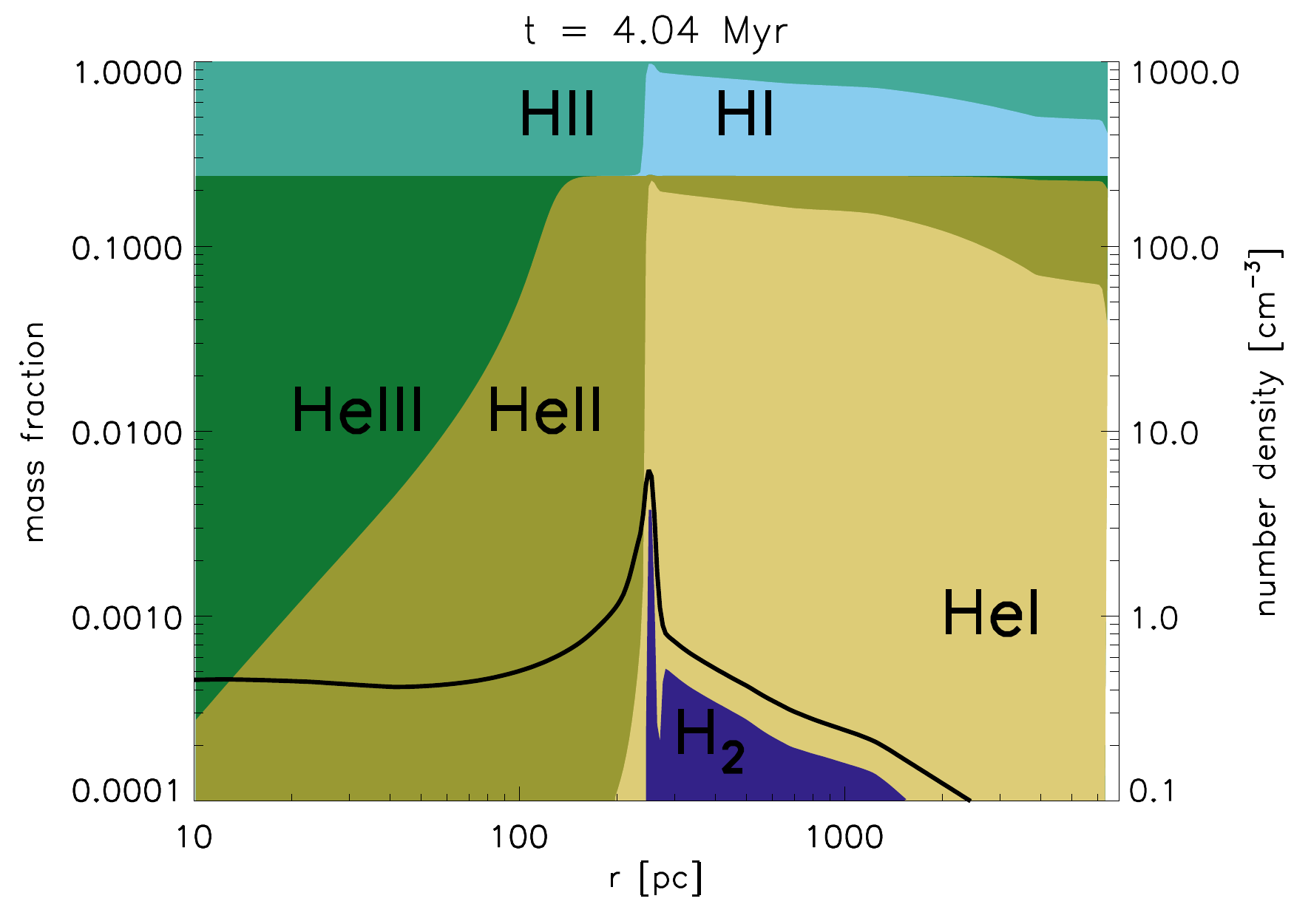}
  \includegraphics[width=0.99\columnwidth]{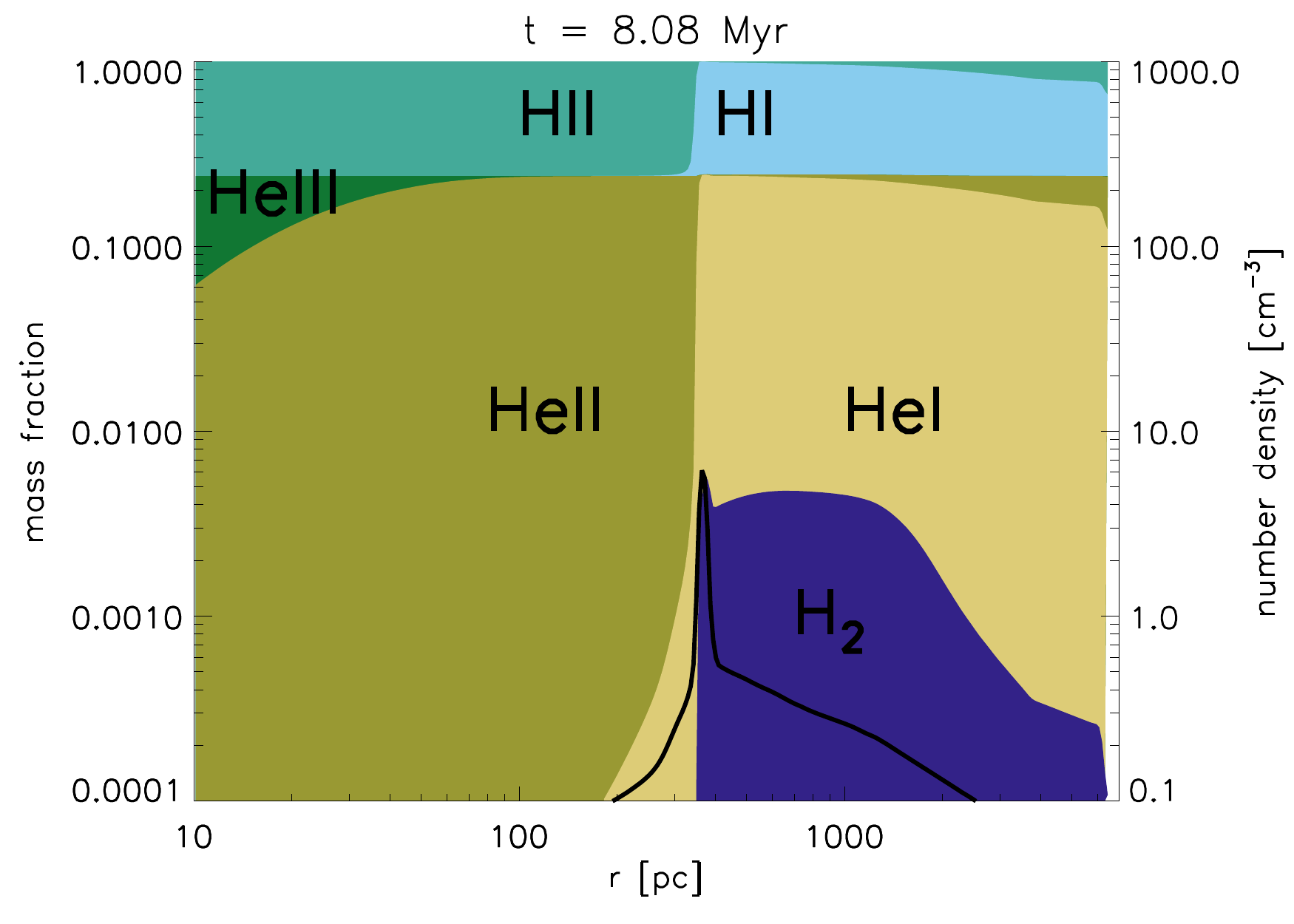}
  \includegraphics[width=0.99\columnwidth]{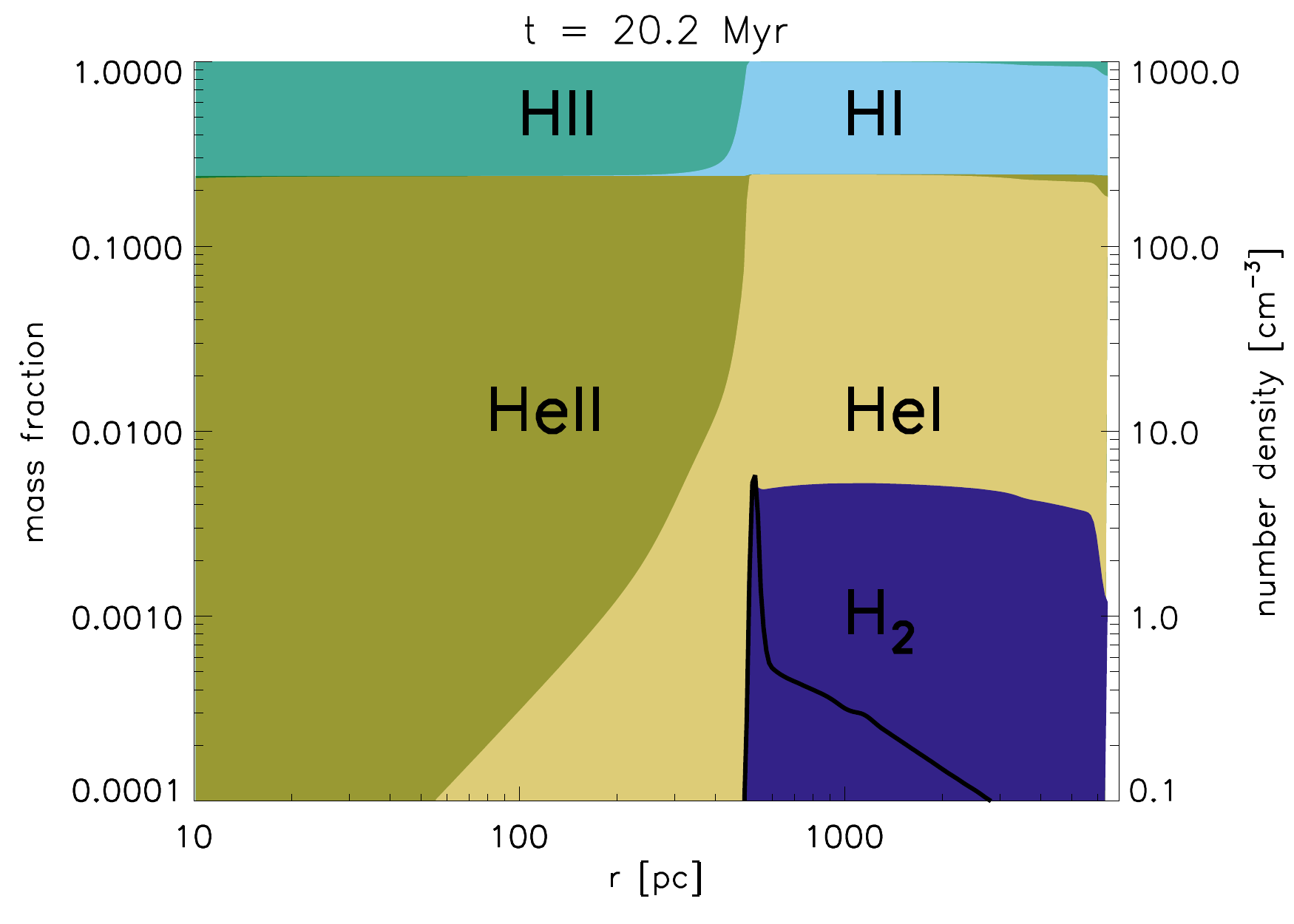}
\caption{Abundances of the primordial species as a function of radius for six output times for a 
1.0\% SFE with a flat IMF in Halo B.} 
\label{fig:abund-blluis100}
\end{figure*}
Similar to the case just mentioned, here, the ionisation front quickly 
advances through the halo, leading to a sudden increase in the LW near-field escape 
fraction from 0 to 100\% at the beginning of the simulation. 
However, after some time, the radiation source weakens and produces fewer ionising photons. 
As a result, these haloes quickly recombine 
in an outside-in fashion, triggering the production of molecular hydrogen. 
The near-field escape fraction drops to 0\% when recombination sets in. 

This can be clearly observed in the lower limit SFE case of a flat IMF in Halo B. 
In Figure~\ref{fig:ifront-blluis100}, a fast outbreak of the I-front yields 
a LW escape fraction of 100\% almost immediately after the simulation starts. 
The abundances and number density are displayed in Figure~\ref{fig:abund-blluis100}. 
At 2.6~Myr (upper left panel), the halo is still completely ionised, but as the spectrum weakens, at 
3.0~Myr (upper right panel), the outer part of the halo partly recombines. 
One can observe that at 3.4~Myr (middle left panel), the recombination proceeds outside-in. 
At 4.0~Myr (middle right panel), the I-front has moved back to the shock-front
position. Here, the high density and the large fraction of free electrons produce the 
ideal environment for the formation of molecular hydrogen. The H$_2$ abundance peaks 
close to 0.01 and hinders all LW photons from escaping in the near-field. 
Molecular hydrogen continues to form outside of the shock-front of the halo at later times 
(8.1~Myr, lower left panel) until the end of the simulation (at 20.2~Myr, lower right panel). 
The time-averaged near-field escape fraction is only 16\%, as after the outside-in recombination 
in the halo, LW photons are blocked completely. The LW near-field escape fraction can be 
approximated in simulations of returning I-fronts (indicated by the superscript $^{\mathrm{(r)}}$)as 
\begin{equation}
f_{\mathrm{esc,LW}}^{\mathrm{(r)}} = \frac{t_{\mathrm{return}}}{t_{\mathrm{total}}}, 
\end{equation}
where $t_{\mathrm{return}}$ is the point in time when the I-front starts moving backwards 
through the halo and $t_{\mathrm{total}}$ is the total runtime of the simulation. 
\subsubsection{Slowly moving ionisation front}
\begin{figure}
\includegraphics[width=0.99\columnwidth]{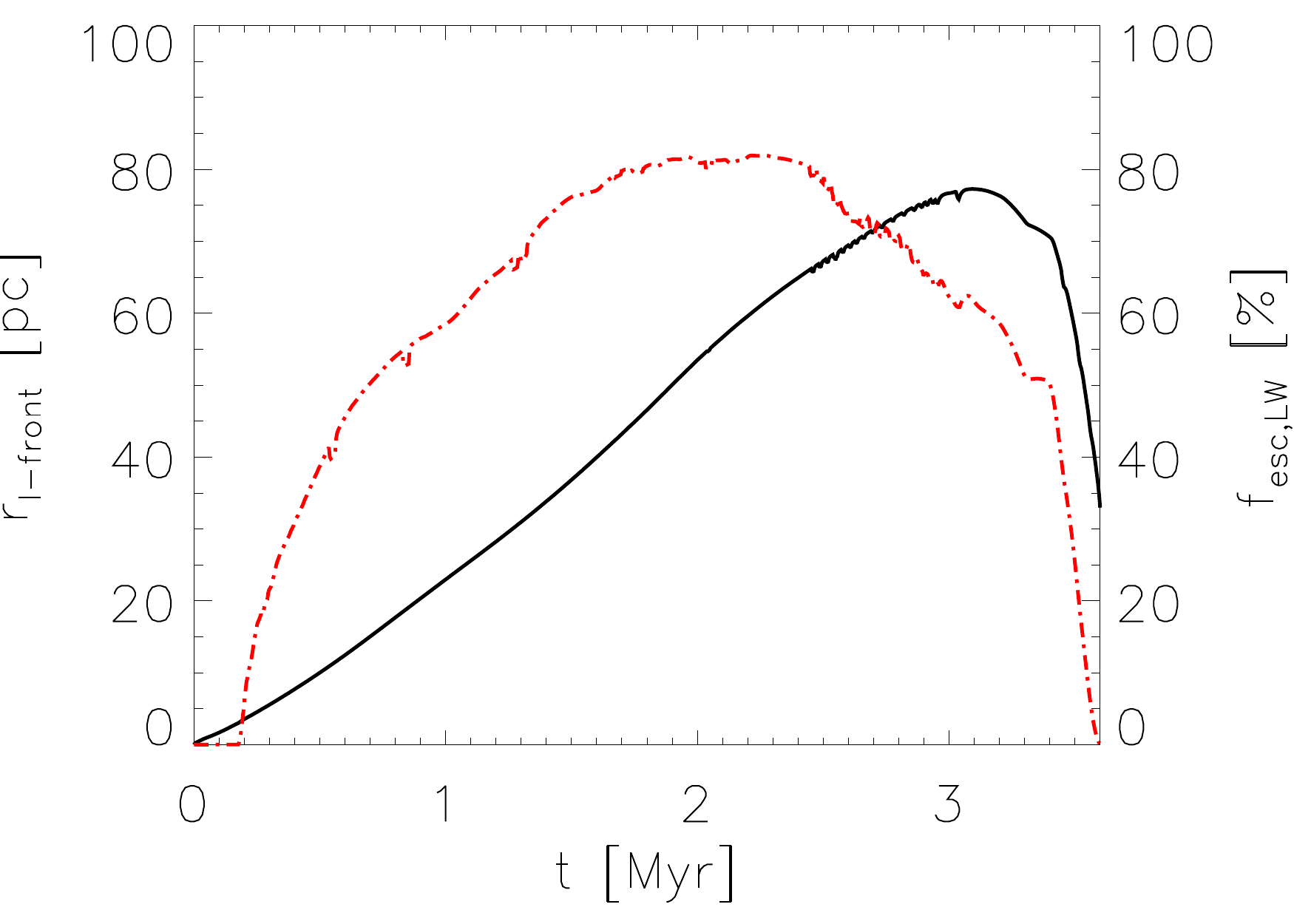}
\caption{Position of the I-front (black solid line) and LW escape fraction in the near-field 
(red dot-dashed line) as a function of time for a 0.1\% SFE with Salpeter IMF in Halo A. }
\label{fig:ifront-ausal010}
\end{figure}
\begin{figure*}
  \includegraphics[width=0.99\columnwidth]{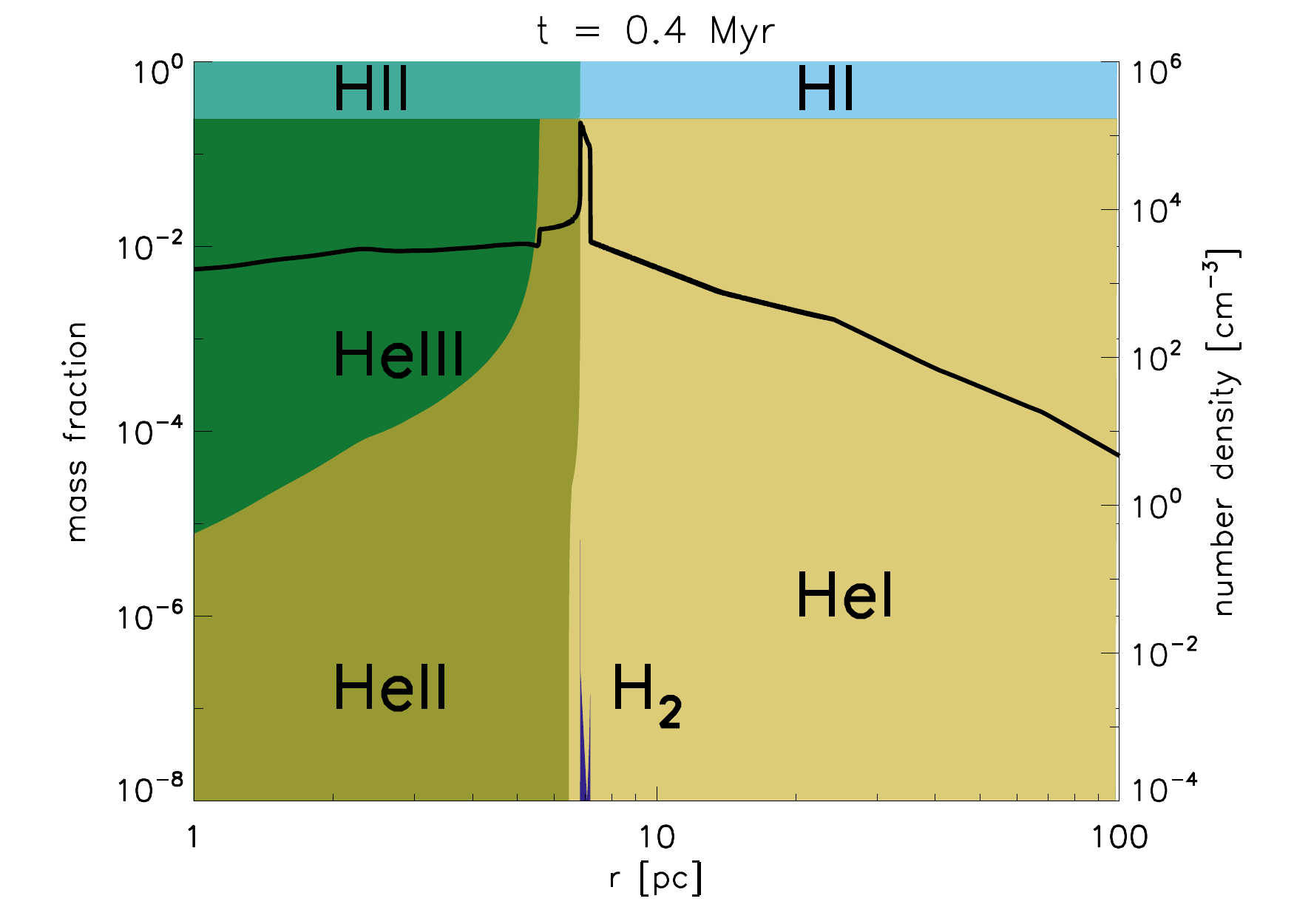}
  \includegraphics[width=0.99\columnwidth]{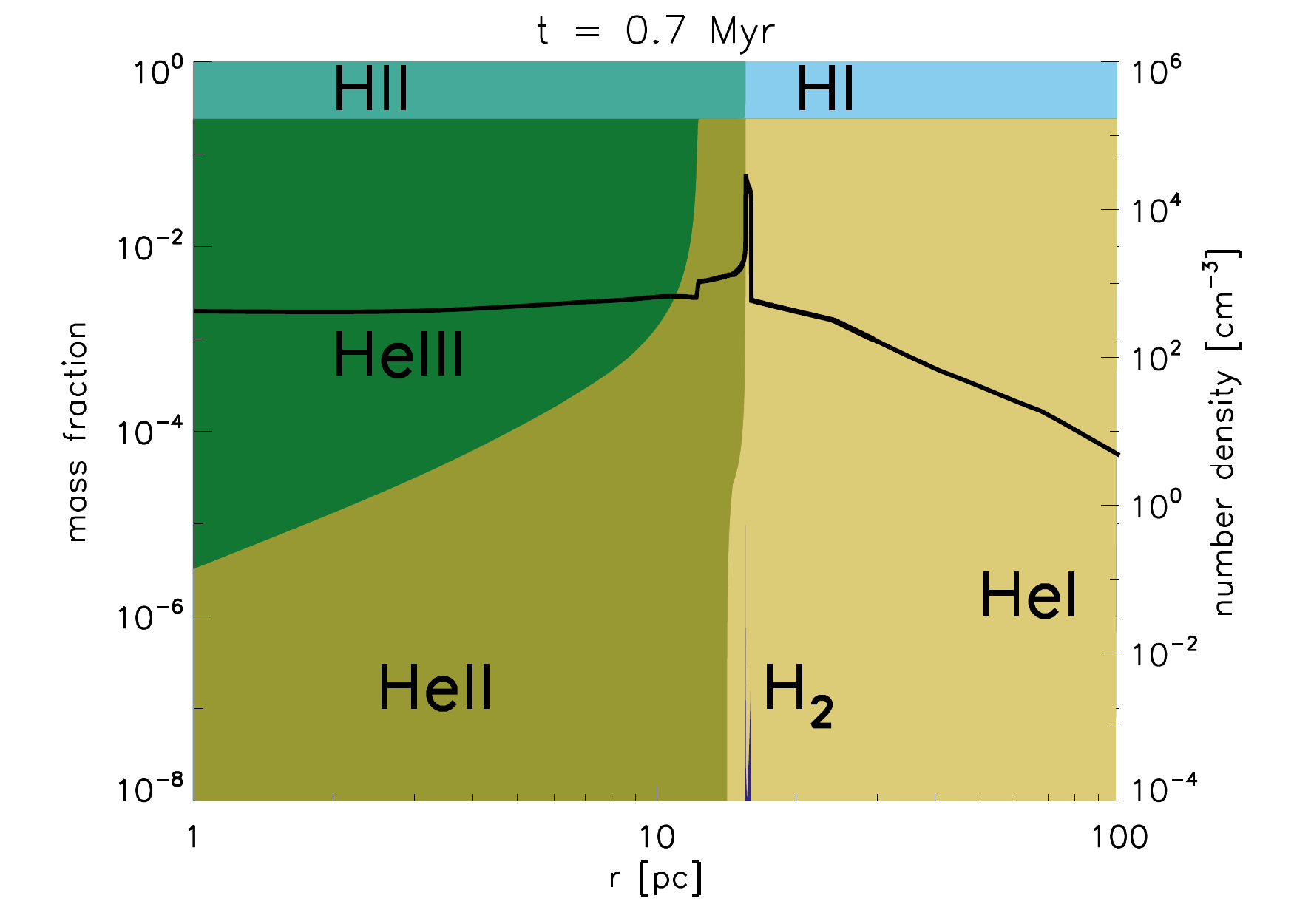}
  \includegraphics[width=0.99\columnwidth]{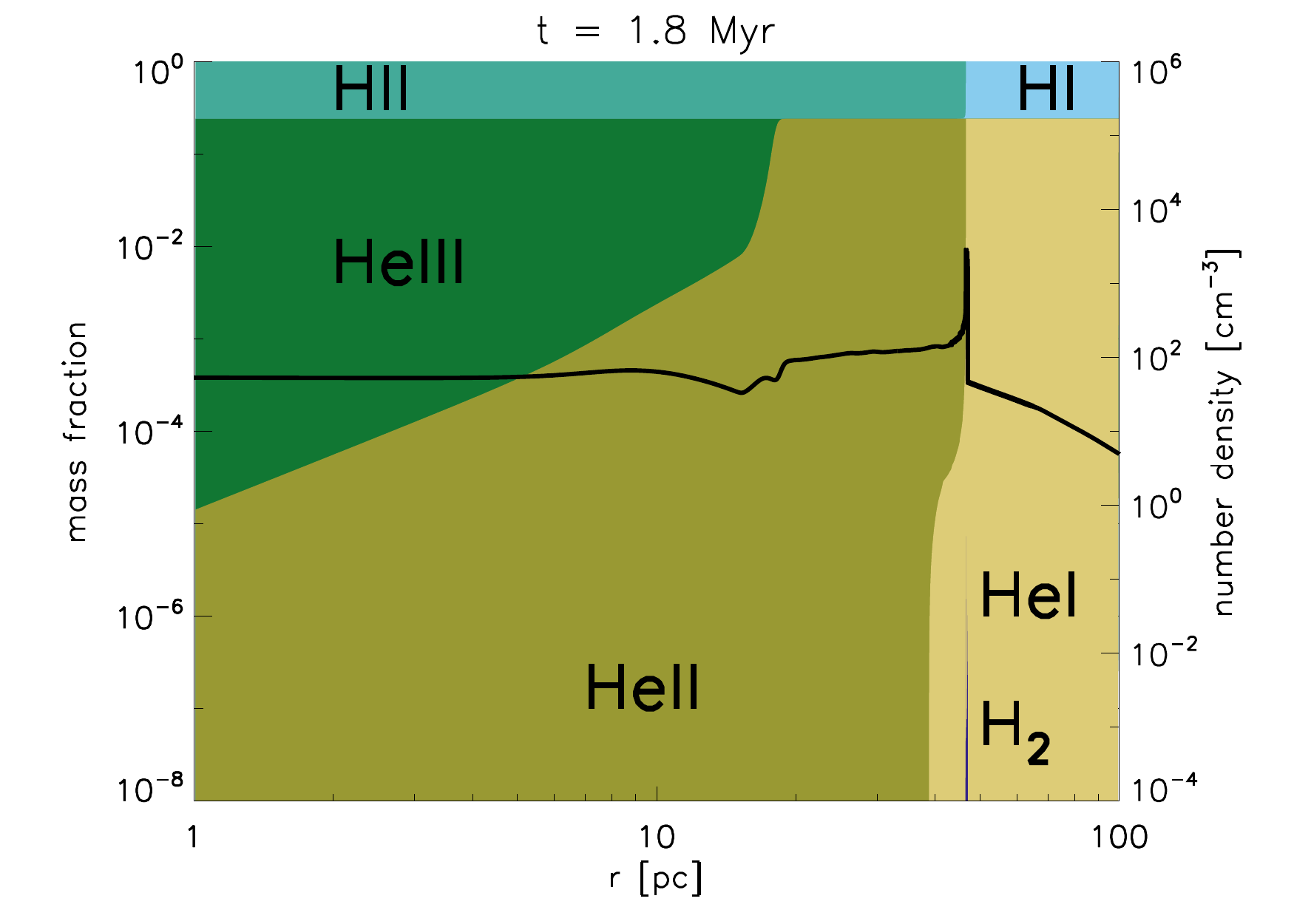}
  \includegraphics[width=0.99\columnwidth]{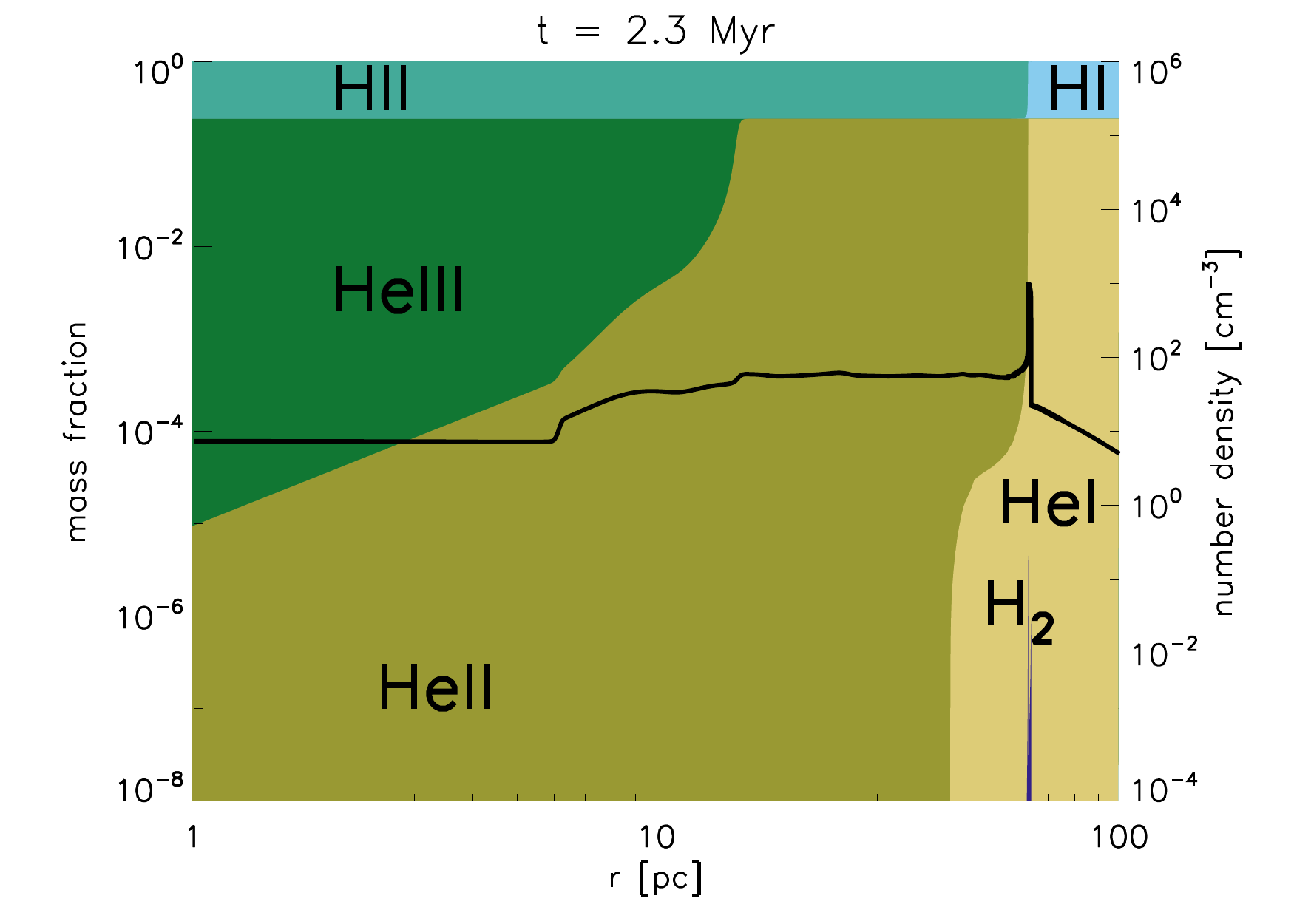}
  \includegraphics[width=0.99\columnwidth]{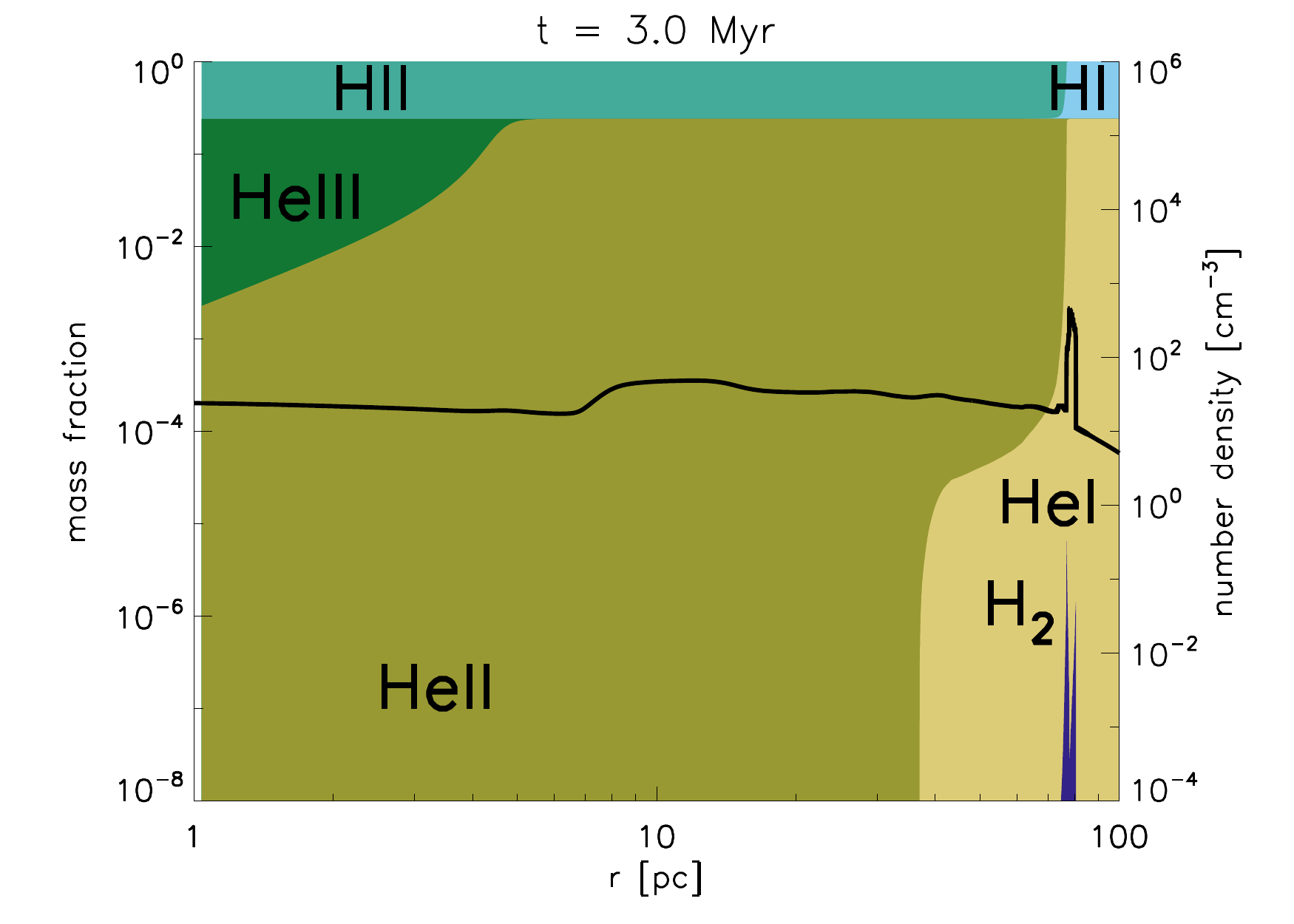}
  \includegraphics[width=0.99\columnwidth]{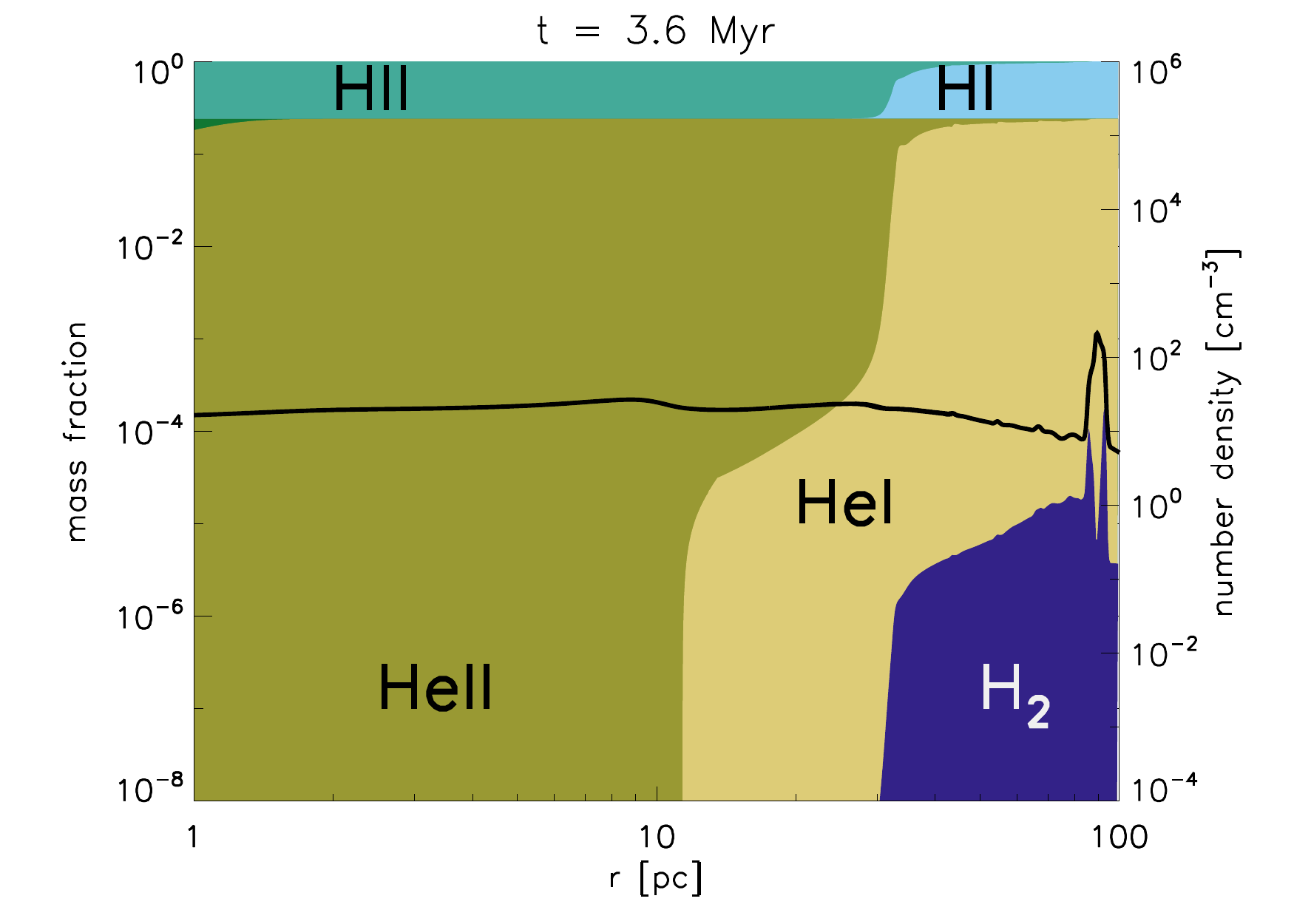}
\caption{Abundances of the primordial species as a function of radius for six output times for a 
0.1\% SFE with a Salpeter IMF in Halo A.} 
\label{fig:abund-ausal010}
\end{figure*}
In the case of a weak radiation source, a D-type I-front advances slowly 
through the halo. 
At its border, the many free electrons trigger massive 
production of H$_2$ and the dense thin shell that builds up reaches optical depths that 
can shield the LW radiation (see also \citealt{rgs01}). Our calculations 
depend critically on the thickness 
and peak abundance of the H$_2$ shell. 
This requires specific consideration and we have to implement a region of high 
numerical resolution at the position of the I-front. 
We adopt a challenging technique where we resolve a 20~pc wide region 
that is bracketing and moving with the I-front with a resolution of 0.001~pc. 
In order to keep the computational time manageable, the first 5\% of the simulations
are run with our standard resolution, introducing a maximum error of 5\% in the
time-averaged LW escape fraction. 

As an example, in Figure~\ref{fig:ifront-ausal010} we show the I-front position and 
the escape fraction as a function 
of time for a 0.1\% SFE and Salpeter IMF in Halo A.
The initial 5\% of the simulation had to be run without the refinement around 
the I-front and therefore shows a LW near-field escape fraction that is probably too low. 
In the upper left panel of Figure~\ref{fig:abund-ausal010} we show the abundances of all 
species at 0.4~Myr. The I-front and the shock front move together. A 
thin shell of molecular hydrogen with peak abundances reaching above $10^{-6}$ forms 
at the position of the I-front that is able to shield some of the LW radiation. The thin H$_2$ shell 
continues moving with the I-front and shock front (upper right panel and middle left panel). 
From 2.3~Myr on (middle right panel), the spectrum starts to weaken and the molecular
hydrogen shell starts to get stronger, preventing more LW radiation from escaping the halo. 
At 3.0~Myr (lower left panel), the He$^{2+}$ abundance decreases and the I-front starts to turn back.
The shell of molecular hydrogen gets thicker, and at the end of the simulation at 3.6~Myr
(lower right panel), all LW radiation is prevented from escaping the halo 
(in the near-field approximation).   
Averaged over the runtime of the simulation, the LW escape fraction in this example is 
63\%. For a slowly moving I-front, we cannot give a simple approximation, in contrast to the 
outbreaking or returning I-front cases. 

%%%
\subsection{Time-averaged escape fractions in the near-field}
%%%
%
\begin{table*}
\begin{center}
\begin{tabular}{l cc|ccccc c}
 &     & & Salpeter   & flat      & log-normal & Kroupa      & Kroupa + nebula  & shielding\\
 & SFE & & 50--500~\Ms & 9--500~\Ms &  1--500~\Ms & 0.1--100~\Ms & 0.1--100~\Ms  &   \\
\hline
Halo A & 0.1 & & 63 (s) &  3 (s) & 38 (s) & 58 (s) & 63 (s) & H$_2$ only\\
       & 0.5 & & 93 (r) & 11 (r) & 26 (r) & 88 (s) & 90 (s) & \\
       & 1.0 & & 99 (o) & 13 (r) & 44 (r) & 90 (s) & 92 (s) & \\
\hline
       & 0.1 & & 59 (s) &  3 (s) & 37 (s) & 47 (s) & 51 (s) & H$_2$ and H \\
       & 0.5 & & 93 (r) & 11 (r) & 26 (r) & 84 (s) & 86 (s) & \\
       & 1.0 & & 99 (o) & 13 (r) & 44 (r) & 88 (s) & 89 (s) & \\
\hline
\hline
Halo B & 1.0 & & 100 (o) & 16 (r) & 61 (r) & 66 (r) & 70 (r) & H$_2$ only\\
       & 2.1 & & 100 (o) & 18 (r) & 78 (r) & 86 (r) & 88 (r) & \\
       & 5.0 & & 100 (o) & 20 (r) & 94 (r) & 96 (r) & 97 (r) & \\
\hline
       & 1.0 & & 99 (o)  & 16 (r) & 61 (r) & 64 (r) & 68 (r) & H$_2$ and H \\
       & 2.1 & & 100 (o) & 18 (r) & 78 (r) & 84 (r) & 86 (r) & \\
       & 5.0 & & 100 (o) & 20 (r) & 94 (r) & 95 (r) & 96 (r) & \\
\end{tabular}
\caption{Time-averaged LW escape fractions in the near-field, averaged over the 
runtime of the simulation. All values are given in percent. 
The upper six rows refer to Halo A, the lower six rows to Halo B. The upper three 
rows of haloes A and B, respectively, are calculated with H$_2$ shielding only, 
the lower three rows with H and H$_2$ shielding. 
The letter in parentheses
refers to the case of I-front behaviour: (o) stands for outbreaking I-front, (r) for 
returning I-front and (s) for slowly moving I-front.}
\label{tab:nearfield}
\end{center} 
\end{table*}
\begin{table*}
\begin{center}
\begin{tabular}{lcc|cccccc}
&     & & Salpeter   & flat      & log-normal & Kroupa      & Kroupa + nebula & shielding\\
& SFE & & 50--500~\Ms & 9--500~\Ms &  1--500~\Ms & 0.1--100~\Ms & 0.1--100~\Ms &    \\
\hline
Halo A & 0.1 & & 54 (s) & 27   (s) & 29 (s) & 41   (s) & 48   (s) & H$_2$ only\\
       & 0.5 & & 87 (r) & 94   (r) & 45 (r) & 57   (s) & 64   (s) & \\
       & 1.0 & & 97 (o) &100   (r) & 85 (r) & 66   (s) & 72   (s) & \\
\hline
       & 0.1 & & 54 (s) & 27   (s) & 28 (s) & 40   (s) & 47   (s) & H$_2$ and H \\
       & 0.5 & & 87 (r) & 93   (r) & 44 (r) & 56   (s) & 63   (s) & \\
       & 1.0 & & 97 (o) & 99   (r) & 84 (r) & 65   (s) & 71   (s) & \\
\hline
\hline
Halo B & 1.0 & &  99 (o) & 100 (r) &  97 (r) & 46 (r) & 58 (r) & H$_2$ only\\
       & 2.1 & & 100 (o) & 100 (r) &  99 (r) & 70 (r) & 76 (r) & \\
       & 5.0 & & 100 (o) & 100 (r) & 100 (r) & 92 (r) & 93 (r) & \\
\hline
       & 1.0 & &  99 (o) &  99 (r) &  96 (r) & 45 (r) & 58 (r) & H$_2$ and H \\
       & 2.1 & &  99 (o) &  99 (r) &  98 (r) & 69 (r) & 75 (r) & \\
       & 5.0 & &  99 (o) &  99 (r) &  99 (r) & 91 (r) & 92 (r) & \\
\end{tabular}
\caption{Time-averaged LW escape fractions in the near-field, averaged over the 
first two Myr. All values are given in percent. 
The structure of the rows is 
as in Table \ref{tab:nearfield}.}
\label{tab:nearfield2Myr}
\end{center} 
\end{table*}
For all SFE, SED and halo combinations, we average the LW escape fraction 
over the runtime of the simulation, 20.2~Myr (or 3.6~Myr for the Salpeter IMF). 
We list the corresponding escape fractions 
in Table \ref{tab:nearfield}. 

Shielding by neutral hydrogen only makes a slight difference: the LW 
escape fractions are at most 16\% higher if we only consider H$_2$ self-shielding 
compared to the case where 
we also account for shielding from H. 
In many cases the difference is negligible. 
This result is in contrast to S15, where we found significantly lower LW 
escape fractions when including shielding by neutral hydrogen. 

As expected, a higher SFE increases the LW escape fraction. This can be observed 
in both haloes and for all SEDs. 
Outbreaking I-fronts lead to LW escape fractions $\ge$ 95\%. 
For slowly moving or returning I-fronts, the values can be as small as 3\%. 
This is mostly observed for the case of a flat SED ranging from 9 to 500~\Ms\ where 
many massive stars die at the beginning of the simulation leading 
to a much weaker ionising source of photons later on. As soon as the more 
massive stars have died, the production of molecular hydrogen sets in 
and the halo becomes optically thick for the majority of the runtime of the
simulations. In all other simulations, the change in spectrum is not 
as extreme and ionising photons are produced for a longer period, 
leading to higher LW escape fractions. 

Our simulations assume that no stars explode as supernovae.
Instead, at the end of their lives, we simply remove their contributions 
from the SED. We made this simplifying assumption for several reasons. 
First, we expect the dynamics of the supernova remnant to be highly sensitive to 
the 3D matter distribution in the protogalaxy (see e.g. \citealt{mml99}). 
It is therefore questionable how well we can represent their effects in 
the simplified geometry of a 
1D model. Second, considerable uncertainties still exist regarding 
the final masses and explosion energies of Pop~III stars. In particular, 
the effects of rotation-driven mass loss may be significant but not 
yet fully understood \citep{emm08}. Finally, the supernovae that explode 
may be unable to clear all of the gas from the protogalaxy, thus resulting in  
a system polluted by metals and dust. 
Our assumption of 
a primordial gas composition may then become highly inaccurate, depending on the 
final metallicity of the gas. 

Nevertheless, once stars begin to explode as supernovae, it is possible 
that they will rapidly drive gas out of the protogalaxy, leading to escape 
fractions quickly reaching 100\%. To account for this possibility, we have 
computed LW near-field escape fractions that are averaged over only the 
first 2~Myr, rather than over the lifetime of the longest-lived massive 
star (Table \ref{tab:nearfield2Myr}).
Two Myr corresponds to the lifetime of stars of 
a few hundred solar masses \citep{schae02} and we can safely assume that no supernovae
exploded earlier. 

Most LW escape fractions averaged over 2~Myr are high; for SFEs with 
a flat IMF in Halo B this behaviour is most pronounced. 
The four outbreaking I-front cases show slightly reduced LW escape fractions, but only 
down to 97\%. 
This change occurs because the
initial few timesteps until outbreak with escape fractions smaller 
than 100\% now carry a slightly larger weight. For slowly moving I-fronts, the 
LW escape fraction decreases (with the exception of a 0.1 SFE flat IMF in Halo A). 
Here, the LW escape fraction slowly increases over time as the I-front moves 
through the halo. 
These last results need to be taken as lower limits as we have to run the slowing moving 
I-front simulations with higher resolution, but can do so only after 1~Myr (0.18~Myr for 
the Salpeter IMF) due to the high computational costs of the required refinement technique.   
LW escape fractions of returning I-fronts decrease or increase depending on 
when the I-front first breaks out (in the case of a Kroupa IMF quite late) and 
starts to return. We therefore cannot give a general trend. 

%%%
\subsection{LW flux-averaged escape fractions in the near-field}
%%%
%
\begin{table*}
\begin{center}
\begin{tabular}{lcc|cccccc}
&     & & Salpeter   & flat      & log-normal & Kroupa      & Kroupa + nebula & shielding\\
& SFE & & 50--500~\Ms & 9--500~\Ms &  1--500~\Ms & 0.1--100~\Ms & 0.1--100~\Ms  &   \\
\hline
Halo A & 0.1 & & 66 & 27 & 50  & 55 & 60 &  H$_2$ and H \\
       & 0.5 & & 92 & 92 & 61  & 82 & 85 & \\
       & 1.0 & & 99 & 99 & 82  & 86 & 88 & \\
\hline
\hline
Halo B & 1.0 & & 100 & 100 & 91 & 74 & 78 & H$_2$ and H \\
       & 2.1 & & 100 & 100 & 96 & 89 & 90 &\\
       & 5.0 & & 100 & 100 & 99 & 97 & 97 &\\
\end{tabular}
\caption{LW flux  averaged LW escape fractions in the near-field, 
averaged over the total LW flux of the simulation. All values are given in percent. 
The upper three rows refer to Halo A, the lower three rows refer to Halo B. 
Shielding by H$_2$ only and by the combination of H and H$_2$ yields the same 
results. }
\label{tab:lwflux}
\end{center} 
\end{table*}
\begin{figure}
\includegraphics[width=0.99\columnwidth]{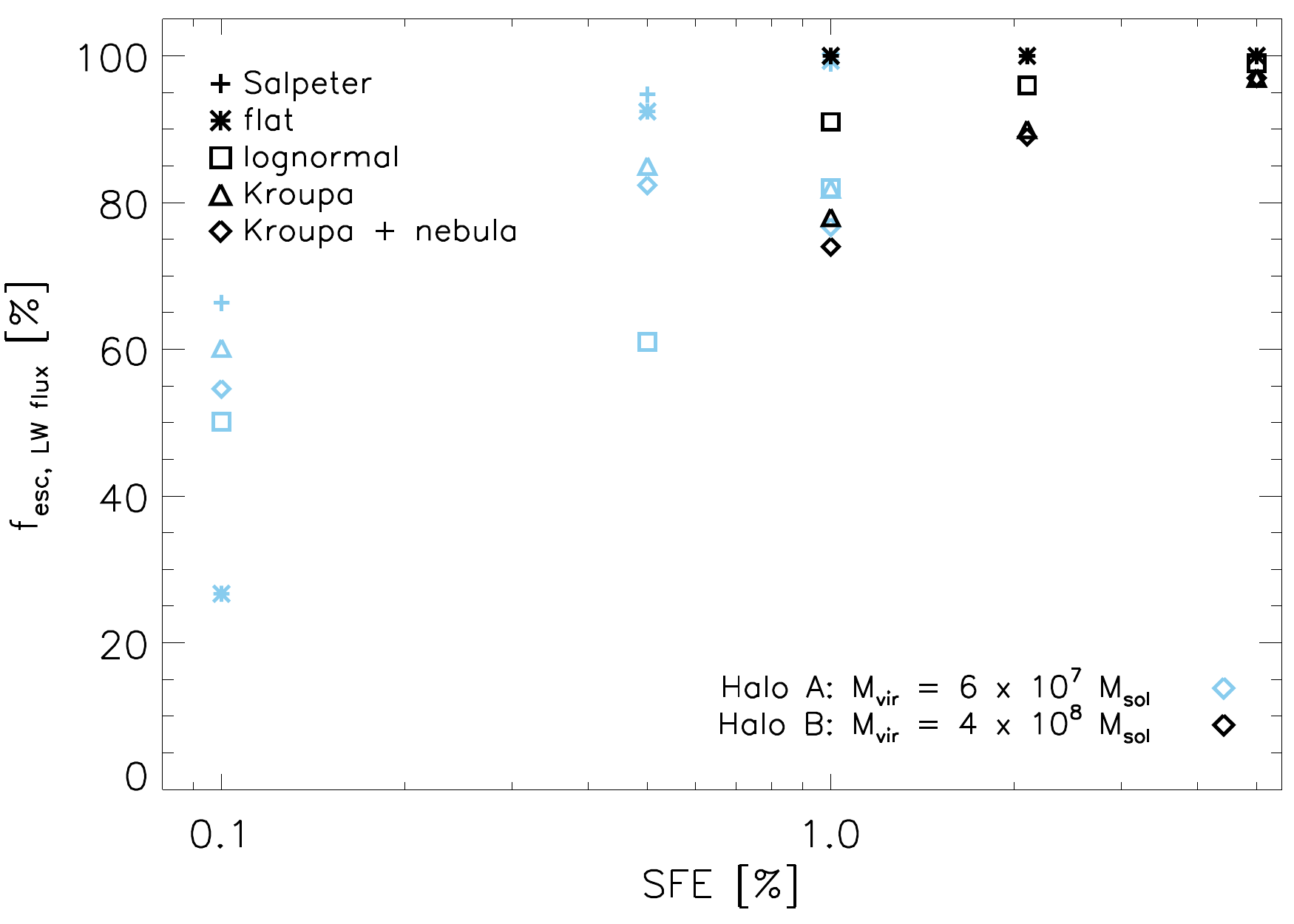}
\caption{Flux-averaged LW escape fractions in the near-field for our parameter space. 
Different SEDs are shown by different symbols, Halo A is colour-coded in light blue, 
Halo B in black. }
\label{fig:lwflux}
\end{figure}
The LW flux of the SED is decreasing over time (compare Figure~\ref{fig:photoncounts}) 
by factors of up to $4\times 10^4$ in the case of the flat IMF. 
We therefore provide the LW escape fraction weighted by the LW flux emitted 
by the stellar source rather than averaged over time. 
We list these values in Table \ref{tab:lwflux} and show them in Figure \ref{fig:lwflux}. 
Except for a SFE of 0.1\%, the escape fraction is always larger than 
75\%.
The difference to the time-averaged LW escape fraction from 
Section 3.2 is most significant for the flat IMF, as in that case, the LW escape fraction 
drops when the LW flux drops and the overall contribution is small. 
Additional shielding by atomic hydrogen does not decrease the LW escape fraction 
in this limit by more than 0.5\%. 
Therefore we only provide the escape fraction obtained by considering 
both H$_2$ and H shielding together.
%%% 
\subsection{Time-averaged escape fractions in the far-field}
%%%
\begin{table*}
\begin{center}
\begin{tabular}{lcc|cccccc}
&    & & Salpeter   & flat      & log-normal & Kroupa      & Kroupa + nebula  & \\
& SFE & & 50--500~\Ms & 9--500~\Ms &  1--500~\Ms & 0.1--100~\Ms & 0.1--100~\Ms  &   \\
\hline
Halo A & 0.1 & &  98 & 73  & 96  & 99  & 100 &  H$_2$ only\\
       & 0.5 & &  99 & 85  & 96  & 100 & 100 &\\
       & 1.0 & & 100 & 84  & 98  & 100 & 100 &\\
\hline
       & 0.1 & & 79  & 61  & 83  & 68  & 68  & H$_2$ and H \\
       & 0.5 & & 91  & 78  & 88  & 84  & 84  &\\
       & 1.0 & & 94  & 79  & 90  & 86  & 86  &\\
\hline
\hline
Halo B & 1.0 & & 100 & 85  & 99  &  99 & 100 &  H$_2$ only\\
       & 2.1 & & 100 & 86  & 100 & 100 & 100 &\\
       & 5.0 & & 100 & 88  & 100 & 100 & 100 &\\
\hline
       & 1.0 & & 97  & 76  & 91  & 87  & 87  &  H$_2$ and H \\
       & 2.1 & & 98  & 78  & 93  & 90  & 90  &\\
       & 5.0 & & 99  & 80  & 95  & 93  & 93  &\\
\end{tabular}
\caption{Time-averaged LW escape fractions in the far-field. All values are given in percent. 
The upper six rows refer to Halo A, the lower six rows to Halo B. The upper three 
rows of haloes A and B, respectively, are calculated with H$_2$ shielding only, 
the lower three rows with H and H$_2$ shielding. }
\label{tab:farfield}
\end{center} 
\end{table*}
In the far-field, the LW escape fractions are generally larger than in the near-field, 
varying from 59\% to 100\%. Our values are listed in Table \ref{tab:farfield}.
Including shielding by neutral hydrogen can reduce the LW escape fraction by 
up to 29\% (comparing 0.1\% SFE with a Kroupa SED in Halo A) and thus should not 
be forgotten in simulations. For the fiducial SFE, all LW escape fractions 
are higher than 75\% and we conclude that the host haloes of the first galaxies do not 
suppress the build-up of the LW background by a large factor. 
%%% 
\subsection{Lower limits for ionising escape fractions}
%%%
Since we track the position of the I-front through 
the halo, we are able to calculate the escape fractions of ionising 
photons as well. The values need to be taken as a first order approximation only, 
since our one dimensional setup prevents us from seeing the  
outbreak of ionised cones like authors recently do with high-resolution
three dimensional studies 
\citep{ritt12,jp15fesc,stacy16}. 

\begin{table*}
\begin{center}
\begin{tabular}{lcc|ccccc}
&    & & Salpeter   & flat      & log-normal & Kroupa      & Kroupa + nebula \\
& SFE & & 50--500~\Ms & 9--500~\Ms &  1--500~\Ms & 0.1--100~\Ms & 0.1--100~\Ms     \\
\hline
Halo A & 0.1 & &  0 &  0 &  0 & 0  & 0 \\
       & 0.5 & & 85 & 10 & 18 & 0  & 0 \\
       & 1.0 & & 96 & 13 & 36 & 0  & 0 \\
\hline
\hline
Halo B & 1.0 & & 98  & 17 & 54 & 26 & 26 \\
       & 2.1 & & 99  & 19 & 66 & 42 & 42 \\
       & 5.0 & & 100 & 22 & 84 & 58 & 58 \\
\end{tabular}
\caption{Time-averaged ionising escape fractions in the near-field. All values are given in percent. 
The upper three rows refer to Halo A, the lower three rows to Halo B. 
All values are derived from the position of the I-front and provide lower limits. }
\label{tab:fesc-ion}
\end{center} 
\end{table*}
The escape fraction of ionising photons is determined by the position 
of the I-front alone. For each simulation-output, we check if the 
I-front has crossed the virial radius, in which case we take 
$f_\mathrm{esc,ion}(t) =$~1; otherwise, $f_\mathrm{esc,ion}(t) =$~0. 
Averaging these values over the runtime of the simulation yields 
the ionising escape fractions  $f_\mathrm{esc,ion}$, listed 
in Table \ref{tab:fesc-ion}. 

The values of  $f_\mathrm{esc,ion}$ are similar to $f_\mathrm{esc,LW}$ in the near-field
for the case of an outbreaking or a returning I-front, since the LW 
escape fractions depend critically on the position of the I-front. For the
case of a slowly moving I-front that never crosses the virial 
radius, we obtain ionising escape fractions of zero while the LW 
escape fractions vary from 3\% to 85\% in the near-field. In these cases, the 
outer halo is optically thick to ionising photons, but optically thin to
LW photons. 
%
% ***************************************************************************
\section{Discussion and Conclusion}
% ***************************************************************************
We studied a multidimensional grid of parameters to calculate the LW escape fractions 
in both the near-field and the far-field limit. We consider two haloes of $5.6 \times 10^7$ 
and $4.0 \times 10^8$~\Ms, that are irradiated by a stellar population in their centre. 
Different SEDs, ranging from a Salpeter IMF between 50 and 500~\Ms\  to a Kroupa 
IMF between 0.1 and 100~\Ms, combined with SFEs between 0.1\% and 5.0\% are taken 
into account. 

We find that the near-field LW escape fraction depends on the motion of the I-front and we
therefore group our results into three different ways the I-front can behave: 
\begin{itemize} 
\item For an outbreaking I-front, the LW escape fraction in the near-field is 
$f_\mathrm{esc}^{\mathrm{(o)}}\ge$ 95\%. 
This is obtained for SFEs greater than 0.5\% with a Salpeter slope IMF ranging 
from 50 to 500~\Ms: the I-front overruns the halo and all gas is ionised for 
the rest of the simulation. 
\item In the case of a spectrum that weakens substantially over time,
haloes recombine outside-in. Large numbers of free electrons still present 
in the outer parts of the halo lead to the aforementioned molecular hydrogen 
formation in the outer part of the halo. The halo becomes optically thick. The 
LW escape fraction can be approximated by 
$f_\mathrm{esc}^{\mathrm{(r)}} \approx t_\mathrm{return} / t_\mathrm{total}$, 
where $t_\mathrm{return}$ is the time when the recombination sets in 
and $t_\mathrm{total}$ is the runtime of the simulation. 
\item For a slowly moving I-front, the LW escape fraction can be as small 
as 3\% and as large as 88\%, 
depending on a thin shell of molecular hydrogen that builds up at the position 
of the I-front. In this case, the stellar radiation acts as an agent of negative feedback 
that enhances the abundance of molecular hydrogen and suppresses the LW background 
(see also \citealt{rgs01}, discussing H$_2$-shell formation in the IGM). 
\end{itemize}
Out of 30 simulations, we find 4 overrunning, 17 returning 
and 9 slowly moving I-fronts. 

In the far-field, all LW escape fractions are much larger than in the near-field, 
as we are shifting out of the line centres and average the LW escape fraction 
over the whole LW energy range. Here, shielding by neutral hydrogen starts to play 
a role since Lyman lines of H$\beta$ and higher, that lie inside the LW range, 
become very broad \citep{har00}. Neutral hydrogen is able to reduce the 
LW escape fraction in the far-field 
by up to 29\%. In general, all LW escape fractions in the far-field are higher than 
75\% in the fiducial case and the halo only slightly hinders the build-up 
of the LW background. 

We also tabulated values of the ionising escape fractions for our parameter space, 
depending on the position of the I-front relative to the virial radius. 
For slowly moving I-fronts, this value equals zero, since the I-front does not reach out 
to the virial radius at any time in our simulation. For simulations where the I-front 
quickly breaks out of the halo, the ionising escape fractions are larger than 95\%. 
If the I-front first breaks out and later returns due to outside-in recombination 
of the halo, the escape fraction ranges from 10\% to 85\%.   

The biggest drawback of this study is our 1D approach. In 3D simulations, 
ionising photons break cones into the haloes, following low-density regions. 
We expect our LW escape fractions to be close to 1 in these directions, 
leading to an overall higher $f_\mathrm{esc, LW}$. We therefore consider our 
LW escape fractions to be lower limits. 

Additionally, our UV source is placed in the centre of our 1D halo and 
all relative velocities of the contributing stars are neglected. 
For the slowly moving I-front cases, the LW near-field escape fractions 
are underestimated in this work. For outbreaking or returning I-fronts, 
the abundance of molecular hydrogen is either too low or too high to 
be influenced by relative motions at the centre of the halo. 

We have shown in this study that the LW escape fraction from the first galaxies
is high. 
Unless a protogalaxy harbours a star cluster with flat IMF or forms stars with a 
low $\sim 0.1$\% SFE, a large fraction of its LW radiation can escape from the host halo.
The LW escape fractions in the far-field exceeds 60\% in all cases. We thus conclude 
that the gas present in the host halo of a stellar cluster with 
$10^4$~\Ms\ 
to a few times $10^6$~\Ms\ in stars cannot prevent the build-up of the LW background. 

The recently observed Ly$\alpha$ system CR7 at $z=6.6$ \citep{cr7}
is comprised of two star-forming components and one seemingly metal-poor
component \citep[however see][]{bowler16}. \cite{bhaskarcr7} showed that the
star-forming components can easily produce the LW radiation field required for DCBH
formation in the metal-poor clump. For an exponentially declining star formation
history, they find that DCBH formation can occur in similar systems as early as
$z\sim 20$, if the star-forming component consists of a DM halo or haloes of
$M_\mathrm{DM} \sim 5 \times 10^8$~\Ms\ and a SFE of $8\%$ or larger (Agarwal et al. in preparation).
This closely matches our case of Halo B forming stars at its upper limit of SFE$\sim5\%$ with the same underlying IMF and stellar mass cut--offs. Given the separation between the star-forming component
and the metal-poor component of CR7 over which the LW feedback allows for DCBH formation (Agarwal et al. 2016),
we can apply our near-field limit. Thus we can conclude that the star-forming component of CR7 type systems could have an escape fraction as high as 96\% at onset of DCBH formation.

In general, the near-field can be applied to neighbouring systems that move with no or
only a small velocity relative to the halo emitting LW radiation. The critical relative
velocity separating the two regimes depends on the strength of the LW absorption lines in
the emitting halo. When the lines are weak and dominated by thermal broadening, $v_{\rm
crit} \simeq v_{\rm th, H_{2}}$, the thermal velocity of the H$_{2}$, which in our
simulated haloes is typically 7--15~km$\,$s$^{-1}$. On the other hand, if the main LW lines are strong
and dominated by Lorentz broadening (i.e.\ if the near-field LW escape fraction is very
small), then the linewidths and hence the critical relative velocity can both be much
larger.

\section*{Acknowledgments}
We would like to thank Mordecai-Marc Mac Low, Mattis Magg and Massimo Ricotti for useful discussions. 
Furthermore, we would like to thank the anonymous referee for his / her valuable comments.  
A.\,T.\,P.\,S., B.\,A., C.-E.\,R. and D.\,J.\,W. acknowledge support 
from the European Research Council under the European Community's Seventh Framework 
Programme (FP7/2007 - 2013) via the ERC Advanced Grant ``STARLIGHT: Formation of the 
First Stars" (project number 339177). 
B.\,A. acknowledges support of a TCAN postdoctoral fellowship at Yale.
S.\,C.\,O.\,G. and R.\,S.\,K. also acknowledge support from 
the Deutsche Forschungsgemeinschaft via SFB 881, ``The Milky Way System'' (sub-projects
B1, B2 and B8) and SPP 1573 , ``Physics of the Interstellar Medium'' (grant number GL 668/2-1).
M.\,L. acknowledges funding from European Union's Horizon 2020 research and innovation 
programme under the Marie Sklodowska-Curie grant agreement No 656428.
E.\,Z. acknowledges funding from the Swedish Research Council (project number 2011-5349).
All simulations were run on the Sciama Supercomputer at the Institute
of Cosmology and Gravitation at the University of Portsmouth. 

\bibliographystyle{mn2e}
\setlength{\bibhang}{2.0em}
\setlength\labelwidth{0.0em}
\bibliography{refs}

\begin{thebibliography}{66}
\expandafter\ifx\csname natexlab\endcsname\relax\def\natexlab#1{#1}\fi

\bibitem[{{Abgrall} {et~al}\mbox{.}(1992){Abgrall}, {Le Bourlot}, {Pineau Des
  Forets}, {Roueff}, {Flower}, \& {Heck}}]{abg92}
{Abgrall} H., {Le Bourlot} J., {Pineau Des Forets} G., {Roueff} E., {Flower}
  D.~R., {Heck} L., 1992, \aap, 253, 525

\bibitem[{{Abgrall} \& {Roueff}(1989)}]{ar89}
{Abgrall} H., {Roueff} E., 1989, \aaps, 79, 313

\bibitem[{{Agarwal} {et~al}\mbox{.}(2016){Agarwal}, {Johnson}, {Zackrisson},
  {Labbe}, {van den Bosch}, {Natarajan}, \& {Khochfar}}]{bhaskarcr7}
{Agarwal} B., {Johnson} J.~L., {Zackrisson} E., {Labbe} I., {van den Bosch}
  F.~C., {Natarajan} P., {Khochfar} S., 2016, \mnras, 460, 4003

\bibitem[{{Agarwal} {et~al}\mbox{.}(2012){Agarwal}, {Khochfar}, {Johnson},
  {Neistein}, {Dalla Vecchia}, \& {Livio}}]{agarw12}
{Agarwal} B., {Khochfar} S., {Johnson} J.~L., {Neistein} E., {Dalla Vecchia}
  C., {Livio} M., 2012, \mnras, 425, 2854

\bibitem[{{Alexander} \& {Natarajan}(2014)}]{at14}
{Alexander} T., {Natarajan} P., 2014, Science, 345, 1330

\bibitem[{{Alvarez}, {Wise} \& {Abel}(2009){Alvarez}, {Wise}, \&
  {Abel}}]{awa09}
{Alvarez} M.~A., {Wise} J.~H., {Abel} T., 2009, \apjl, 701, L133

\bibitem[{{Anninos} {et~al}\mbox{.}(1997){Anninos}, {Zhang}, {Abel}, \&
  {Norman}}]{anet97}
{Anninos} P., {Zhang} Y., {Abel} T., {Norman} M.~L., 1997, New Astronomy, 2,
  209

\bibitem[{{Begelman}, {Volonteri} \& {Rees}(2006){Begelman}, {Volonteri}, \&
  {Rees}}]{begel06}
{Begelman} M.~C., {Volonteri} M., {Rees} M.~J., 2006, \mnras, 370, 289

\bibitem[{{Bowler} {et~al}\mbox{.}(2016){Bowler}, {McLure}, {Dunlop}, {McLeod},
  {Stanway}, {Eldridge}, \& {Jarvis}}]{bowler16}
{Bowler} R.~A.~A., {McLure} R.~J., {Dunlop} J.~S., {McLeod} D.~J., {Stanway}
  E.~R., {Eldridge} J.~J., {Jarvis} M.~J., 2016, ArXiv 1609.00727

\bibitem[{{Bromm} \& {Loeb}(2003)}]{bl03}
{Bromm} V., {Loeb} A., 2003, \apj, 596, 34

\bibitem[{{Clark}, {Glover} \& {Klessen}(2008){Clark}, {Glover}, \&
  {Klessen}}]{cgk08}
{Clark} P.~C., {Glover} S.~C.~O., {Klessen} R.~S., 2008, \apj, 672, 757

\bibitem[{{Clark} {et~al}\mbox{.}(2011){Clark}, {Glover}, {Klessen}, \&
  {Bromm}}]{cgkb11}
{Clark} P.~C., {Glover} S.~C.~O., {Klessen} R.~S., {Bromm} V., 2011, \apj, 727,
  110

\bibitem[{{Dijkstra}, {Ferrara} \& {Mesinger}(2014){Dijkstra}, {Ferrara}, \&
  {Mesinger}}]{dfm14}
{Dijkstra} M., {Ferrara} A., {Mesinger} A., 2014, \mnras, 442, 2036

\bibitem[{{Dijkstra}, {Gronke} \& {Sobral}(2016){Dijkstra}, {Gronke}, \&
  {Sobral}}]{markcr7}
{Dijkstra} M., {Gronke} M., {Sobral} D., 2016, \apj, 823, 74

\bibitem[{{Draine} \& {Bertoldi}(1996)}]{db96}
{Draine} B.~T., {Bertoldi} F., 1996, \apj, 468, 269

\bibitem[{{Eisenstein} \& {Loeb}(1995)}]{el95}
{Eisenstein} D.~J., {Loeb} A., 1995, \apj, 443, 11

\bibitem[{{Ekstr{\"o}m}, {Meynet} \& {Maeder}(2008){Ekstr{\"o}m}, {Meynet}, \&
  {Maeder}}]{emm08}
{Ekstr{\"o}m} S., {Meynet} G., {Maeder} A., 2008, in IAU Symposium, Vol. 250,
  Massive Stars as Cosmic Engines, {Bresolin} F., {Crowther} P.~A., {Puls} J.,
  eds., pp. 209--216

\bibitem[{{Ferland} {et~al}\mbox{.}(2013){Ferland}, {Porter}, {van Hoof},
  {Williams}, {Abel}, {Lykins}, {Shaw}, {Henney}, \& {Stancil}}]{cloudy13}
{Ferland} G.~J. {et~al.}, 2013, \rmxaa, 49, 137

\bibitem[{{Fryer}, {Woosley} \& {Heger}(2001){Fryer}, {Woosley}, \&
  {Heger}}]{fwh01}
{Fryer} C.~L., {Woosley} S.~E., {Heger} A., 2001, \apj, 550, 372

\bibitem[{{Haiman}, {Abel} \& {Rees}(2000){Haiman}, {Abel}, \& {Rees}}]{har00}
{Haiman} Z., {Abel} T., {Rees} M.~J., 2000, \apj, 534, 11

\bibitem[{{Hartwig} {et~al}\mbox{.}(2016){Hartwig}, {Latif}, {Magg}, {Bromm},
  {Klessen}, {Glover}, {Whalen}, {Pellegrini}, \& {Volonteri}}]{tilmancr7}
{Hartwig} T. {et~al.}, 2016, \mnras, 462, 2184

\bibitem[{{Hirano} {et~al}\mbox{.}(2014){Hirano}, {Hosokawa}, {Yoshida},
  {Umeda}, {Omukai}, {Chiaki}, \& {Yorke}}]{hir14}
{Hirano} S., {Hosokawa} T., {Yoshida} N., {Umeda} H., {Omukai} K., {Chiaki} G.,
  {Yorke} H.~W., 2014, \apj, 781, 60

\bibitem[{{Inayoshi}, {Haiman} \& {Ostriker}(2016){Inayoshi}, {Haiman}, \&
  {Ostriker}}]{iho16}
{Inayoshi} K., {Haiman} Z., {Ostriker} J.~P., 2016, \mnras, 459, 3738

\bibitem[{{Jarosik} {et~al}\mbox{.}(2011){Jarosik}, {Bennett}, {Dunkley},
  {Gold}, {Greason}, {Halpern}, {Hill}, {Hinshaw}, {Kogut}, {Komatsu},
  {Larson}, {Limon}, {Meyer}, {Nolta}, {Odegard}, {Page}, {Smith}, {Spergel},
  {Tucker}, {Weiland}, {Wollack}, \& {Wright}}]{wmap7b}
{Jarosik} N. {et~al.}, 2011, \apjs, 192, 14

\bibitem[{{Kitayama} {et~al}\mbox{.}(2004){Kitayama}, {Yoshida}, {Susa}, \&
  {Umemura}}]{ket04}
{Kitayama} T., {Yoshida} N., {Susa} H., {Umemura} M., 2004, \apj, 613, 631

\bibitem[{{Koushiappas}, {Bullock} \& {Dekel}(2004){Koushiappas}, {Bullock}, \&
  {Dekel}}]{kbd04}
{Koushiappas} S.~M., {Bullock} J.~S., {Dekel} A., 2004, \mnras, 354, 292

\bibitem[{{Kroupa}(2001)}]{kroupa01}
{Kroupa} P., 2001, \mnras, 322, 231

\bibitem[{{Latif} {et~al}\mbox{.}(2013){Latif}, {Schleicher}, {Schmidt}, \&
  {Niemeyer}}]{latif13c}
{Latif} M.~A., {Schleicher} D.~R.~G., {Schmidt} W., {Niemeyer} J., 2013,
  \mnras, 433, 1607

\bibitem[{{Latif} \& {Volonteri}(2015)}]{lv15}
{Latif} M.~A., {Volonteri} M., 2015, \mnras, 452, 1026

\bibitem[{{Leitherer} {et~al}\mbox{.}(1999){Leitherer}, {Schaerer}, {Goldader},
  {Delgado}, {Robert}, {Kune}, {de Mello}, {Devost}, \&
  {Heckman}}]{starburst99}
{Leitherer} C. {et~al.}, 1999, \apjs, 123, 3

\bibitem[{{Lodato} \& {Natarajan}(2006)}]{ln06}
{Lodato} G., {Natarajan} P., 2006, \mnras, 371, 1813

\bibitem[{{Loeb} \& {Rasio}(1994)}]{lr94}
{Loeb} A., {Rasio} F.~A., 1994, \apj, 432, 52

\bibitem[{{Mac Low} \& {Ferrara}(1999)}]{mml99}
{Mac Low} M.-M., {Ferrara} A., 1999, \apj, 513, 142

\bibitem[{{Madau} \& {Rees}(2001)}]{mad01}
{Madau} P., {Rees} M.~J., 2001, \apjl, 551, L27

\bibitem[{{Marigo} {et~al}\mbox{.}(2001){Marigo}, {Girardi}, {Chiosi}, \&
  {Wood}}]{mar01}
{Marigo} P., {Girardi} L., {Chiosi} C., {Wood} P.~R., 2001, \aap, 371, 152

\bibitem[{{Mas-Ribas}, {Dijkstra} \& {Forero-Romero}(2016){Mas-Ribas},
  {Dijkstra}, \& {Forero-Romero}}]{lluis16}
{Mas-Ribas} L., {Dijkstra} M., {Forero-Romero} J.~E., 2016, \apj, 833, 65

\bibitem[{{Mortlock} {et~al}\mbox{.}(2011){Mortlock}, {Warren}, {Venemans},
  {Patel}, {Hewett}, {McMahon}, {Simpson}, {Theuns}, {Gonz{\'a}les-Solares},
  {Adamson}, {Dye}, {Hambly}, {Hirst}, {Irwin}, {Kuiper}, {Lawrence}, \&
  {R{\"o}ttgering}}]{mort11}
{Mortlock} D.~J. {et~al.}, 2011, \nat, 474, 616

\bibitem[{{Omukai}(2001)}]{O01}
{Omukai} K., 2001, \apj, 546, 635

\bibitem[{{Omukai}, {Schneider} \& {Haiman}(2008){Omukai}, {Schneider}, \&
  {Haiman}}]{osh08}
{Omukai} K., {Schneider} R., {Haiman} Z., 2008, \apj, 686, 801

\bibitem[{{O'Shea} {et~al}\mbox{.}(2015){O'Shea}, {Wise}, {Xu}, \&
  {Norman}}]{oshea15}
{O'Shea} B.~W., {Wise} J.~H., {Xu} H., {Norman} M.~L., 2015, \apjl, 807, L12

\bibitem[{{Paardekooper}, {Khochfar} \& {Dalla Vecchia}(2015){Paardekooper},
  {Khochfar}, \& {Dalla Vecchia}}]{jp15fesc}
{Paardekooper} J.-P., {Khochfar} S., {Dalla Vecchia} C., 2015, \mnras, 451,
  2544

\bibitem[{{Pallottini} {et~al}\mbox{.}(2015){Pallottini}, {Ferrara}, {Pacucci},
  {Gallerani}, {Salvadori}, {Schneider}, {Schaerer}, {Sobral}, \&
  {Matthee}}]{pcr7-15}
{Pallottini} A. {et~al.}, 2015, \mnras, 453, 2465

\bibitem[{{Park} \& {Ricotti}(2011)}]{pm11}
{Park} K., {Ricotti} M., 2011, \apj, 739, 2

\bibitem[{{Raiter}, {Schaerer} \& {Fosbury}(2010){Raiter}, {Schaerer}, \&
  {Fosbury}}]{raiter10}
{Raiter} A., {Schaerer} D., {Fosbury} R.~A.~E., 2010, \aap, 523, A64

\bibitem[{{Ricotti}, {Gnedin} \& {Shull}(2001){Ricotti}, {Gnedin}, \&
  {Shull}}]{rgs01}
{Ricotti} M., {Gnedin} N.~Y., {Shull} J.~M., 2001, \apj, 560, 580

\bibitem[{{Ritter} {et~al}\mbox{.}(2012){Ritter}, {Safranek-Shrader}, {Gnat},
  {Milosavljevi{\'c}}, \& {Bromm}}]{ritt12}
{Ritter} J.~S., {Safranek-Shrader} C., {Gnat} O., {Milosavljevi{\'c}} M.,
  {Bromm} V., 2012, \apj, 761, 56

\bibitem[{{Rodgers} \& {Williams}(1974)}]{rw74}
{Rodgers} C.~D., {Williams} A.~P., 1974, \jqsrt, 14, 319

\bibitem[{{Rydberg} {et~al}\mbox{.}(2015){Rydberg}, {Zackrisson}, {Zitrin},
  {Guaita}, {Melinder}, {Asadi}, {Gonzalez}, {{\"O}stlin}, \&
  {Str{\"o}m}}]{ry15}
{Rydberg} C.-E. {et~al.}, 2015, \apj, 804, 13

\bibitem[{{Schaerer}(2002)}]{schae02}
{Schaerer} D., 2002, \aap, 382, 28

\bibitem[{{Schauer} {et~al}\mbox{.}(2015){Schauer}, {Whalen}, {Glover}, \&
  {Klessen}}]{anna15}
{Schauer} A.~T.~P., {Whalen} D.~J., {Glover} S.~C.~O., {Klessen} R.~S., 2015,
  \mnras, 454, 2441

\bibitem[{{Shang}, {Bryan} \& {Haiman}(2010){Shang}, {Bryan}, \&
  {Haiman}}]{sbh10}
{Shang} C., {Bryan} G.~L., {Haiman} Z., 2010, \mnras, 402, 1249

\bibitem[{{Sobral} {et~al}\mbox{.}(2015){Sobral}, {Matthee}, {Darvish},
  {Schaerer}, {Mobasher}, {R{\"o}ttgering}, {Santos}, \& {Hemmati}}]{cr7}
{Sobral} D., {Matthee} J., {Darvish} B., {Schaerer} D., {Mobasher} B.,
  {R{\"o}ttgering} H.~J.~A., {Santos} S., {Hemmati} S., 2015, \apj, 808, 139

\bibitem[{{Stacy} \& {Bromm}(2014)}]{sb14}
{Stacy} A., {Bromm} V., 2014, \apj, 785, 73

\bibitem[{{Stacy}, {Bromm} \& {Lee}(2016){Stacy}, {Bromm}, \& {Lee}}]{stacy16}
{Stacy} A., {Bromm} V., {Lee} A.~T., 2016, \mnras, 462, 1307

\bibitem[{{Stacy}, {Greif} \& {Bromm}(2010){Stacy}, {Greif}, \&
  {Bromm}}]{stacy10}
{Stacy} A., {Greif} T.~H., {Bromm} V., 2010, \mnras, 403, 45

\bibitem[{{Tumlinson}(2006)}]{tum06}
{Tumlinson} J., 2006, \apj, 641, 1

\bibitem[{{Volonteri}(2010)}]{vol10}
{Volonteri} M., 2010, \aapr, 18, 279

\bibitem[{{Whalen}, {Abel} \& {Norman}(2004){Whalen}, {Abel}, \&
  {Norman}}]{wan04}
{Whalen} D., {Abel} T., {Norman} M.~L., 2004, \apj, 610, 14

\bibitem[{{Whalen} \& {Norman}(2006)}]{wn06}
{Whalen} D., {Norman} M.~L., 2006, \apjs, 162, 281

\bibitem[{{Whalen} \& {Norman}(2008)}]{wn08b}
{Whalen} D., {Norman} M.~L., 2008, \apj, 673, 664

\bibitem[{{Whalen} \& {Fryer}(2012)}]{wf12}
{Whalen} D.~J., {Fryer} C.~L., 2012, \apjl, 756, L19

\bibitem[{{Wise} \& {Fuhr}(2009)}]{wf09}
{Wise} W.~L., {Fuhr} J.~R., 2009, \jpcrd, 38, 565

\bibitem[{{Wolcott-Green} \& {Haiman}(2011)}]{wh11}
{Wolcott-Green} J., {Haiman} Z., 2011, \mnras, 412, 2603

\bibitem[{{Wolcott-Green}, {Haiman} \& {Bryan}(2011){Wolcott-Green}, {Haiman},
  \& {Bryan}}]{whb11}
{Wolcott-Green} J., {Haiman} Z., {Bryan} G.~L., 2011, \mnras, 418, 838

\bibitem[{{Wu} {et~al}\mbox{.}(2015){Wu}, {Wang}, {Fan}, {Yi}, {Zuo}, {Bian},
  {Jiang}, {McGreer}, {Wang}, {Yang}, {Yang}, {Thompson}, \& {Beletsky}}]{wu15}
{Wu} X.-B. {et~al.}, 2015, \nat, 518, 512

\bibitem[{{Zackrisson} {et~al}\mbox{.}(2011){Zackrisson}, {Rydberg},
  {Schaerer}, {{\"O}stlin}, \& {Tuli}}]{zack11}
{Zackrisson} E., {Rydberg} C.-E., {Schaerer} D., {{\"O}stlin} G., {Tuli} M.,
  2011, \apj, 740, 13

\end{thebibliography}

\label{lastpage}

\end{document}